# Nanophotonics and DNA: New approaches

Vasil G. Bregadze[1], Zaza G. Melikishvili[2], Tamar G. Giorgadze[1]

[1]Department of Biological System Physics,

Javakhishvili Tbilisi State University, Andronikashvili Institute of Physics, Tbilisi, Georgia

[2]Department of Coherent Optics and Electronics,

Georgian Technical University, Chavchanidze Institute of Cybernetics, Tbilisi, Georgia

Email: vbregadze@gmail.com, v.bregadze@aiphysics.ge

**Abstract**

The aim of the present work is spectroscopic and thermodynamic study of DNA catalytic properties in the following processes: a) redox; b) formation of inter–strand cross-links; c) performing of photo-dynamic effects; d) nanoscale resonance radiationless electron excitation energy transfer. The most attention is paid to the latter, as truly nanoscale method in its origin.

The nanoscale method of laser induced fluorescence resonance energy transfer (FRET) to donor (acridine orange) - acceptor (ethidium bromide) intercalator pair for quantitative and qualitative study of stability quality DNA double helix in solution in real time is offered.

FRET method allows to estimate the concentration of double helix areas with high quality stability applicable for intercalation in DNA after it was subjected to stress effect. It gives the opportunity to compare various types of DNAs with 1) different origin; 2) various damage degrees; 3) being in various functional state.

Alternative model and mechanisms of photodynamic effect on DNA in solutions are proposed. They are based on photoenergy degradation in solutions. The energy activates electrolytic dissociation of water molecules on $H_3O^+$ and $OH^-$ and acts as a catalyst for hydrolyse reactions of phosphordiester and glycoside linkages.

**Keywords:** DNA; nanophotonics; photodynamic effect; resonance energy transfer; nanoscale fluorescent probing of stressed DNA.

**Abbreviations:**

AA – Ascorbic acid

AgNPs – Silver nanoparticles

AO – Acridine orange

*bp* – Base pair

CCD – Charge-coupled device

DNA – Calf thymus DNA

DPT – Double proton transfer

EB – Ethidium bromide

$e^{-}_{aq}$ – Hydrated electrons

FRET – Fluorescence resonance energy transfer

IRET – Induced resonance energy transfer

IR – Infrared

$K_{e/aq}$ – Rate constant of the reaction of $e^{-}_{aq}$

K – Rate constant

$M^{n+}$ – metal ion

NPs – Nanoparticles

## 1 Introduction

Bionanophotonics is the recently emerged, but already well defined, truly interdisciplinary field of science and technology aimed at establishing and using the peculiar properties of light and nanoscale light–matter interaction with an emphasis on life science applications [1-9].

The deoxyribonucleic acid (DNA) molecule is well known as a blueprint of life, amazingly rich in information content and very robust. However, its unique structural features and powerful recognition capabilities can also be of interest for assembling artificial structures for a variety of applications in nanophotonics. A DNA helix is itself a nanoobject that can be manipulated in various ways, but it can also be treated as a versatile molecular scaffold for building nanoscale devices from the bottom up [1]. DNA's unique properties are intensively studied from different points of view: as a molecular wire, as a drug delivery system, a ladder for ordered arrangements of various nanostructures, a spacer to control distances between nanoobjects, etc.

There is tendency in modern medicine lately to use such nanoscale interaction, as so-called Fluorescence Resonance Energy Transfer (FRET) in donor-acceptor pair of dye molecules which intensely absorb light in the visible regions of spectrum and have significant quantum yield [10-15]. Microscopic FRET is also used for cytological diagnosis of tumors [16-19].

On the other hand, there are nearly no works where quantitative estimation of FRET applied to DNA is given. For instance, energy transfer (ET) effectiveness, quantitative estimation of DNA double helix conditions, accessibility of nuclear DNA to intercalator, etc. have not been yet estimated. In this connection it is up-to-day the modeling of DNA defects in solution and estimation of the quality of double helix. It is interesting to study a stress impact on the DNA by transitive metal (TM) ions, laser irradiation and heating.

The aim of the present work is DNA double helix ability application to: 1) double proton transfer in GC and AT pairs; 2) mobile absorption of $H_3O^+$ ions; 3) inner mobility including intercalation; 4) formation of densely packed regular structure suitable for light re-emission for spectroscopic and thermodynamic study of DNA catalytic properties in the following processes: a) redox; b) nanoscale resonance radiationless electron excitation energy transfer; c) formation of inter–strand cross-links; d) performing of photo-dynamic effects; e) photoinduced conformation transfers in silver nanoparticles and making of structures having plasmon characteristics.

## 1.1 Inner Mobility of DNA

Comprehensive data on the inner mobility of DNA are collected in [20]. Table 1 presents some characteristics of inner mobility of DNA and RNA. By comparing these movements with their thermodynamic characteristics a certain correlation can be observed. We can predict that for the excitation of small amplitude of atom mobility application of thermal energy is sufficient, but $B \rightarrow A$ and $B \rightarrow Z$ transition excitation requires the change of surrounding (humidity, ionic strength, alcohols, transition metal ions, etc.). Thus for modeling DNA structural changes, stimulated by the interactions with various ligands, particularly, metal ions ($M^{2+}$) it is necessary to estimate the correlation between the energy of interaction and the life time of such complexes on the one hand, and dynamic characteristics of DNA, on the other.

## 1.2 DNA polyelectrolyte characteristics

Electrostatic potential (EP) is an additional property of metal ions, metal atoms and nanoparticles at interactions with DNA. It depends on the composition and sequence of nucleotides and the concentration of counterions. For instance, the sequence polarity of G – C pairs arrangement is – – + ($N_7.O_6 \cdots HN_4$) and for C – G pair it is + – – ($N_4H \cdots O_6.N_7$). The pairs A – T ($N_7HN_6 \cdots O_4$) and T –A ($O_4 \cdots HN_6.N_7$) have similar arrangement – + – [23]. Quantum chemical evaluations show [24, 25] identical sequence for electron-donor atoms arrangement both for the unscreened B – DNA form and for the one screened by $Na^+$ ions:

$$N_7G – N_3G – N_3A – O_2T – O_6G – N_7A$$

At first sight DNA macromolecules are highly charged poly-anions ($2e^-$ charge per a pair of its base). Actually, according to Manning [26], at ionic strength $10^{-2}$ about 76% of phosphate groups are in the complex with $Na^+$ ions which means that they are in neutral state. It, in its turn, facilitates interaction of transition metal ions, atoms and small nanoparticles with electron – donor atoms present in major grooves of DNA double helix. On the other hand, negatively charged phosphate groups and counter ions (mobile ions) lead to formation of hydrodynamic layers near the DNA surface. Thus the DNA sites with condensed $Na^+$ ions and the sites surrounded by hydrodynamic layers (water and mobile ions) will be in dynamic equilibrium. So transition metal ions such as Mn(II), Co(II), Ni(II), Zn(II) and Cu(II) not only form outer-sphere complexes with DNA but also form outer inter chain cross-links with DNA double helix. At the same time Cu(I) and Ag(I) ions form inner sphere complexes with DNA particularly chelate $N_7G$ –$O_6G$ complex and inter- strand $N_1G$-M(I)-$N_3C$ cross-links. Special attention should be paid to $H^+$ because $H^+$ ions are exception and they are characterized by mobile – adsorption state on DNA surface. It is demonstrated by: 1) Abnormally strong effect of $H^+$ ions on DNA spectra in UV region, and strong influence of ionic strength on the interaction between $H^+$ and DNA [27]; 2) Weak effect of $H^+$ ions on DNA stability in the complex with intercalators; 3) Absence of AgNPs oxidation in DNA complexes by $H^+$ ions in the presence of $Ag^+$ or Cu(II) ions; 4) Abnormally high mobility of $H^+$ ions (life time $10^{-13}$ sec in water and $10^{-14}$ sec in ice [28]).

### 1.3 Intercalation

One more interesting characteristics of DNA is intercalation. Intercalator binding coefficients with natural DNA and model nucleotides is between $10^4$–$10^6$ sequence. Such kind of molecule intercalator binding with DNA double helix takes place, as a rule, in two steps. The first step is adsorption of intercalators on periphery areas of the surface; on the second step the intercalators penetrate into the DNA molecule which is accompanied by unfolding and elongation of helix and, certainly, by increasing the helix rigidity. The first step is characterized by weak binding. Binding constants given above are typical for intercalation processes. Besides, the intercalation process is characterized by the principal of excluding the nearest binding places. In case of proflavine only 44% of possible binding places are busy [29]. It means that every second place is vacant. But there are some data that allow to doubt in the statement. In particular, our group has managed to obtain fluorescence spectra of DNA ternary complex DNA–AO–EB at specific concentration of 3 intercalator molecules to 5 base pair of DNA (Figure 1 The authors express their gratitude to Dr . J. Chkhaberidze for the gives spectra [30]). It should be underlined that unlike soft ions such as Ag+, ions of $Ni^{++}$, $Co^{++}$, $Zn^{++}$, $Mn^{++}$, $Cu^{++}$ choose stable nucleotide dimmers. Intercalators at binding with DNA prefer the same dimmers.

For instance, overlapping is much stronger in sequence pyrimidine – 3', 5'– purine than at intercalation of sequence purine – 3', 5' – pyrimidine. It means that in the first case the energy gain is 7–13 kcal/$M^{-1}$ more than in the second case [31].

**1.4 Adsorbing thermodynamic model of DNA interaction with small ligands**

Metal induced catalytic characteristics of DNA are closely related to the modeling of its structural changes caused by interaction with different ligands, such as metal ions, in particular. Thus, for modeling structural changes of DNA, stimulated by the interaction with ligands or metal ions ($M^{2+}$), it is necessary to know the correlation between the energy of the interaction and the lifetime of such complexes on the one hand, and the dynamic characteristics of DNA on the other. It can be concluded that a local influence of transition metal ions on structural changes in DNA is possible only in case when the lifetimes of the complexes are commensurable to the specific times of inner mobility in DNA (oscillation of small groups of atoms, double helix untwisting, opening of the individual base pairs, untwisting of helix binding of proteins and cell division). The movements last from $10^{-10}$ sec to hundreds of seconds and longer. We have managed to find such correlation [32], by applying Frenkel's phenomenological thermodynamic approach [33,34,35], which was used as early as in 1924 for the study of gas adsorption by the surface of solid bodies. Frenkel introduced a new term "lifetime" for adsorption state $\tau$ and connected it to the energy of interaction $|\Delta E|$ between adsorbate and adsorbent surface by the following expression:

$$\tau = \tau_0^{a-s} \cdot \exp(\Delta E / k_B T) \qquad (1)$$

where $\tau_0^{a-s}$ is a single oscillation time of the adsorbate relative to the surface, which is assumed to be $10^{-13}$ – $10^{-12}$ sec; $k_B$ is the Boltzmann constant.

In good approximation to the above model is the study of small ligand interaction with biomacromolecules in solutions. And, indeed, using relation of equilibrium constant $K$ for ligand – biopolymer reactions with the change of Gibbs' free energy $\Delta$G, we get

$$\Delta G = -RT \ln K \, , \qquad (2)$$

where R is gas constant (1,9872 kcal/grad*M), T – absolute temperature per M of substance. We have obtained the final expression for description of time for small ligand interaction with macromolecules

$$\tau = \tau_0 \cdot K \, , \qquad (3)$$

where $\tau$ is lifetime of ligand – macromolecule complexes. Now we must clarify the value $\tau_0$ for solutions. Frenkel's $\tau_0^{a-s}$ is the duration of the fluctuation excitation of adsorbing atoms or molecules interacting with a solid surface , and it is supposed to be equal to the period of the oscillation of adsorbate relative to the adsorbent surface. In solutions the value $\tau_0$ describes the duration of the relaxation of rotary and translation movements of the solvent molecules, ions, solvated ions or low-molecular-weight substances and lies between $10^{-11}$ and $10^{-10}$ sec. Thus, if we assume that *logK=4-6* for DNA binding with twofold positively charged metal ions of the first transition series, and $\tau_0 = 10^{-11}$ sec, then the life span of these complexes is about $10^{-7} – 10^{-5}$ sec.

So, we have managed to correlate stability constant for complex formation *K*, which can be registered in equilibrium state, to dynamic characteristic $\tau$ for the lifetime of a complex. Thus, the principal concept of molecular biophysics regarding biomolecule: structure – dynamics – function can be reformulated as: structure – stability – function. It should be specially noted that such an approach highly simplifies and widens time interval (from $10^{-10}$ sec to $10^5 – 10^6$ sec and more) under investigation of dynamic characteristics of macromolecules.

Derivation of formulae (3) and its detailed ground are given in [32].

## 2 DNA as catalyst in redox reactions

### *2.1 Oxydation*

Figure 2 shows the absorption spectra for AgNPs and AgNPs-DNA complexes. Figure 3 presents first derivatives of absorption spectra for AgNPs and AgNPs in the complex with the DNA.

The analysis of Figures 2 and 3 shows that at DNA interaction with AgNPs a short-wave shift of AgNPs absorption band (6 nm) takes place. Besides, there can be observed a 20% hypochromic effect.

It should be pointed out that the values of hypsochromic shift and hypochromic effects depend on certain extend from the life-time of AgNPs. Distinct trend is observed in decrease of these effects connected with time. The short-wave shift points out that at interaction with DNA there is a kind of loosening of interaction between silver atoms inside the AgNPs. Decrease of the intensity of the absorption band is due to partial corrosion of AgNPs in the presence of DNA. At DNA interaction with AgNPs and $H_3O^+$ in water solutions AgNPs-DNA and $H_3O^+$-DNA complexes are formed. $H_3O^+$ ions being in mobile adsorbed state on the surface of DNA [36,32] can form complexes with AgNPs. Nanoparticles in their turn moisture the DNA surface. We have not found any data on $O_2$ and $NO_3^-$ molecules adsorption on DNA surface though high concentrations of $NO^{3-}$ (0.3M) can oxidize AgNPs (see Figure 4 and 5). Photo-irradiation in water also oxidizes AgNPs [37].

As far as in 1980 [38] one of the authors of the present work proposed definition of various ions and molecules affinity to electrons in water solutions as reciprocal to activation energy $\wedge Ga$ of hydrated electron with organic and inorganic substances in water solutions rate constant of which is evaluated as $k_{e-/aq} = 10^{11} \exp(-\wedge Ga/ RT)$ [39]. Taking as an example study of DNA interaction with transition metal ions by absorption spectrometry and equilibrium dialyses we have shown that as a rule the bigger is $k_{e-/aq}$ value for metal ions the more intensive is their interaction with DNA [39].

Table 2 shows rate constants $k_{e-/aq}$ and $1/\wedge Ga$ for reaction with $Ag^+$, $H_3O^+$, $NO_3^-$ ions and $O_2$, $H_2O$ and DNA molecules, where $\wedge Ga$ is hydrated electron activation energy according to $k_{e-/aq} = 10^{11} \exp(-\wedge Ga/ RT)$. Besides, in the last column of Table 2 concentration of the substances under investigation is given.

It can be explained by the ability of DNA to adsorb $H_3O^+$ (pK~4) and metal ions of the first transition row $M^{2+}$ (pK=4–6) [40]. It should be noted that at study of the interaction of transition metal ions with DNA non-buffer solutions are usually used. As a rule pH of the solutions is 5.5–6. At these pH values the concentration of $H_3O^+$ in the solution is satisfactory for the attacks at DNA because $H_3O^+$ adsorption on DNA is of mobile character [36]. Thus, the surface of DNA can adsorb both AgNPs and $H_3O^+$ and serve as a catalyst in AgNPs oxidation reaction. So, we can state with definite certainty that in AgNPs–$H_3O^+$ complexes, existing on DNA surface, electron transfer from silver atom to $H_3O^+$ can occur because the value of electro negativity of hydrogen atom is bigger than that of silver atom and much more than of $H_3O^+$ ion. We can express the said by the following equation:

$$2AgNPs\ (n) + 2H^+ \rightarrow 2AgNPs\ (n\text{-}1) + (2AgNPs + DNA) + H_2$$

It is necessary to underline that AgNPs used in the experiment have the size of 1-2 nm and, consequently, their concentration is at least two orders less than the one for atoms which constitute the particles. Assuming that all nanoparticles of the solution are adsorbed on DNA and they interact from the side of double helix major groove and their average size is 1.5 nm, we can conclude that the average distance between AgNPs adsorbed on DNA is nearly 60 nm (1 particle for about 170 base pairs). As soon as $H_3O^+$ reduction starts (as a result of AgNPs oxidation) the water in the solution dissociates to $OH^-$ and $H^+$ supplying new $H_3O^+$ until new dynamic balance of the system is achieved. $Ag^+$ ions which are created during the reaction have high stability constant with DNA (pK≥10.8 [36]). As far as in 1969 Wilhelm and Daune [41] showed that $Ag^+$ ions make two kinds of intra-spherical complexes with G-C DNA pairs: chelate $N_7G$ – $O_6G$ and intra-strand linear complex between $N_1G$ and $N_3C$, so-called cross-link. The authors [41] believe that at making the complex of the second type $H_3O^+$ is released from DNA guanine into the solution. It is an additional mechanism of $H_3O^+$ formation in solution. As formation of the second type complex depends on DNA dynamic characteristics, i.e. on frequency of unfolding the base pairs, the possibility of the process will be much higher in case when DNA is denatured. We could observe it in our experiment (see Fig.1).

### *2.2 Reduction*

Figure 6 demonstrates absorption spectra of AgNPs and of the following complexes DNA–$AgNO_3$– AgNPs, DNA–$AgNO_3$–AgNPs–AA. It can be seen that AA reduced silver ions in ternary complex DNA–$AgNO_3$–AgNPs. Thus, AgNPs activate the process of quick reduction of $Ag^+$ ions to silver atoms. Analyzing absorption spectra given on Fig.6 we can draw the following conclusions – 1) $Ag^+$ ions interfere in $H_3O^+$ ion mobility and prevent oxidation of AgNPs and, 2) AgNPs in their turn activate the process of $Ag^+$ ions reduction in the presence of AA. AgNPs absorption spectra shift to red side (6-7nm) and significant increase of absorption spectra intensity undoubtedly point out the increase of AgNPs size in the complex with DNA. It should be underlined that without DNA nothing happens to AgNPs in water solution.

Figure 7 presents AA effect on absorption spectra of DNA complex with silver ions in visible region. By 160 min. we can clearly see appearance of notable spectrum from reduction of silver atoms. In 24 hours absorption spectra specific for silver atoms can be observed. It is seen that the absorption spectra have complex form, they remind absorption spectra of substances having specific resonance interactions. Resonance interactions are specific for structures having identical chromophore and hard structure [42], e.g. for absorption spectra of poly-peptide in α-helix state [43]. Interaction of silver ions, which is characterized by inter-cross type of links with DNA, reduces dynamic characteristics of double helix and makes it harder. The other reason for the complexity of the spectra is reduction of silver ions and their further condensation in clusters and nanoparticals with the size of more than 50 nm [44].

## 3 Photo-induced DNA-dependent Conformational Changes in Silver Nanoparticles

Figure 8 shows AgNPs absorption spectra before and after their heating followed by their cooling. Colloidal silver suspension was placed in hermetic test-tube and it was being incubated in a vessel with boiling water for 15 min. Then the test-tube with the sample was cooled in ice-bath to room temperature. It can be seen from Figure 8 that nano-particles exposed to heating demonstrate noticeable weakening of absorption intensity, and what is very important, widening of absorption spectra from 140 nm to 215 nm.

Hypsochromic shift and widening of AgNPs absorption spectra caused by their interaction with DNA and heating demonstrates, on the one hand, loosening of AgNPs, i.e. penetration of solvent in nanoparticles, the atoms of which show weaker polarizability than water [45,46].

The company NanoComposix [44, 47] gives sample absorption spectra for spherical AgNPs with particle sizes of 10, 20, 30, 40, 50, 60, 70, 80, 90 and 100 nm at the same mass concentration 0.02 mg/ml. The given data show that with the increase of the size of the particles widening of the red shift of absorption band can be observed. It is notable that despite the growth of the particle size, i.e. decrease of their total number in the solution, the intensity of absorption bands for nanoparticles with the sizes from 10 to 40 nm is not practically changed, only a small shift of absorption band maximums can be observed. The above explicitly points out that chromophore units are silver atoms and not nanoparticles. Thus, we can draw a conclusion that silver atoms in AgNPs are sufficiently isolated and bound together by dispersion interaction (induced dipole-induced dipole). As these interactions are performed in water surrounding they should be considerably amplified at the expense of so called hydrophobic effect [48], that means compaction and then minimization of the surface (decrease of the system entropy).

We should especially point out that in nanoparticle, which consists of one kind of atoms, along with the mentioned dispersion interaction, the so-called resonance interaction should take place [42]. Such types of interactions are typical for molecular crystals and they usually lead to exiton splitting of the principal absorption band. Inevitable condition for exciton splitting is the presence of a system consisting of identical groups and having hard structure [43]. The absence of splitting can mean that AgNPs under investigation (Figure 2) have liquid structure resembling a drop which under definite conditions such as temperature, photo-irradiation, variations in dielectric constant of the environment should be characterized by conformational transitions. So, once again analyzing absorption spectra presented in Figure 2 we can make the conclusion that hypsochromic shift shows not only loosening of AgNPs but moreover the transfer of the drop into a spherical segment (wetting angle $\theta < \pi/2$). It means that DNA surface is moistened by silver nanoparticles. First of all it is connected with great affinity of soft ions ($Cu^+$, $Ag^+$, $Hg^{++}$ etc.) and metal ions $M^0$ with DNA double helix [36]. Photo-diffusion of AgNPs (see Figure 10) on DNA double helix resembles flesh-desorption phenomena, i.e. fast heating of AgNPs by photons, then desorption of silver ions with their following adsorption by DNA double helix, including creation of cross-links between silver atoms and DNA chains.

Kinetics of AgNPs photoirradiation has also been studied. Figure 9 and 10 show superposed absorption spectra of free AgNPs and AgNPs in complex with DNA before and after irradiation using water filter.

The analysis of the spectra in Figure 9 and 10 demonstrate that only the irradiated complexes AgNPs-DNA have distinctly expressed isosbestic point. The test with the free AgNPs shows that as a result of photoirradiation desorption of silver atoms and their oxidation to $Ag^+$ ions takes place. The presence of isosbestic points in the absorption spectra of irradiated AgNPs-DNA complexes proves that the system has not less than two states, i.e. AgNPs-DNA complexes have several forms of existence joint by structural photodiffusive transition from one form, e.g. spherical one, to extended long and probably one-dimensional form along DNA double helix. The analysis of the spectra really shows with good correlation (≤ 5%) that the space under the spectra is preserved which means that there are no changes in chromophore electron structure. Besides, half width of absorption spectra $\Delta\lambda_{1/2}$ is changed from 140 nm to 360 mn. Red shift and widening of AgNPs absorption band points out to the typical for molecular systems increase of electron conjugation (linear and cyclic conjugated systems [49]).

We have also carried out evaluation of energy needed for heating of AgNPs (1–2 nm). Under the condition when atomic (specific) heat capacity of liquid silver is equal to 30.5 J/(g-atom × grad); single photon energy ($\lambda$ = 430 nm) is $46.2 \times 10^{-20}$ J it is possible to heat a single silver nanoparticle (size 1nm) consisting of 30 silver atoms up to 610 K; in case when a silver nanoparticle has a size of 2 nm, i.e. consists of 240 atoms, it can be heated up to 340 K. It means that photo-irradiation ($\lambda$ = 430 nm) has absolute ability to cause the photo-diffusion of AgNPs, especially of those with size about 1nm.

To check the concept we have carried out thermodynamic kinetic analysis of absorption spectra of AgNPs-DNA complexes (see Figure 10).

Let's consider the changes in absorption spectra for photo-irradiated AgNPs-DNA complexes given in Figure 10 versus the duration of irradiation in $M_t/M_e$ and $t^{1/2}$ (see Figure 11). $M_e \equiv A_{t=0} - A_{t=6,600}$ is the number of silver atoms in nanoparticles at the beginning (absorption $A_t$ when $\lambda$ = 430 nm at t = 0), $M_t \equiv A_{t=0} - A_t$ is molar quantity of silver atoms desorbed by the time moment t (difference between absorption $A_{t=0} - A_t$ at $\lambda$ = 430 nm). As it can be seen the curves in Figure 11 have S-shape for photo desorption kinetics of silver atoms both from the surface of free AgNPs and AgNPs-DNA complexes. S – shape appearance of the curves denotes that photo-induced desorption of atoms is a complex and multiphase process [37]; it means diffusion of silver atoms from the inner part of a nanoparticle to its surface, conformational changes in the particles especially in those ones that are adsorbed on DNA surface. Next we are going to consider the results given in Figures 9 and 10 in $\ln[M_e/(M_e - M_t)]$ and t coordinates, which are given in Figure 12.

The analysis of the curve shows that only initial stage of the given curves of desorption kinetics obey linear law of Langmuir first-order equation

$$\ln\left[M_e/\left(M_e - M_t\right)\right] = kt. \tag{4}$$

The constant of desorption rate for silver atoms from the surface of AgNPs has been evaluated from the slopes of the curves and the data is: for AgNPs bound with DNA $k_d \cong 9 \times 10^{-5}$ $s^{-1}$. The values allow us to estimate activation energy $E_d$ for desorption reaction using the equation

$$k_d = \upsilon_0 \exp\left(- E_d/RT\right), \tag{5}$$

where $\upsilon_0$ is pre-exponential factor assumed as $\upsilon_0 \approx 10^{10}$ $s^{-1}$ (reciprocal quantity to silver atom oscillation time in nanoparticles). In this case we have acquired the value for $E_d \cong 80$ kJ/M $Ag^0$ for AgNPs bound with DNA at T = 300 K. As $E_d = E_a + Q_a$, where $E_a$ is adsorption activation energy and $Q_a$ is adsorption heat of nanoparticles, so $Q_a \cong 80$ kJ/M $Ag^0$ under the condition that formation of nanoparticles is not an activated process. The value of heat is specific for cluster nanostructures [50].

Silver atom photodesorption from the surface of nanoparticles, included in DNA complex, has desorption activation energy of 80 kJ/M $Ag^0$ and, consequently, we can assume that photo-induced diffusion of AgNPs on DNA double helix takes place along with activation of silver atoms desorption energy which equals to 80 kJ/M $Ag^0$ and thus we can state that the energy of $Ag^0$ interaction with DNA double helix is not less than 80 kJ/M $Ag^0$. In accordance with the Gibbs equation

$$\Delta G = - RT \ln K, \tag{6}$$

and taking into account the evaluated energy 80 kJ/M $Ag^0$, we assume that the stability constant of the complex is not less than $10^{14}$ and, consequently, life time of the complexes is equal to $10^4$ s. Life-times of 1 s order and more are characteristic for inter-strand interactions with the participation of transition metal ions, so called cross-links [22]. As far as in 1969 Wilhelm and Daune [23] showed that $Ag^+$ ions form cross-links between DNA chains thus releasing protons bound with $N_1$ guanine and $N_3$ thymine into the solution. We have estimated stability constants of $Cu^+$ and $Ag^+$ ions with DNA which are equal to pK = 10.8 for $Ag^+$ and pK = 14.9 for $Cu^+$.

Thereafter the change of free energy for $Ag^+$ is 63 kJ/mol and 86 kJ/mol for $Cu^+$ and lifetimes are 0.63 s for $Ag^+$ and $8.6 \times 10^3$ s for $Cu^+$ [4].

## 4 Nonradiative Energy Transfer between Intercalator Molecules and Defects in DNA Duplex

Interaction of intercalator molecules with DNA, particularly, acridine orange (AO) and ethidium bromide (EB) (Fig 14), depends on ionic strength of the solution, DNA nucleotide content, its sequence [51] and double helix structure. Besides, such interaction depends on transition metal ions (TM) [32] which cause ejection of intercalators, though intercalator molecules and TM ions have different binding sites on DNA. We should point out that AO and EB molecules as well as TM ions (Mn(II), Co(II), Ni(II), Cu(II) and Zn(II)) at interaction with DNA double helix have similar values of stability constants and their pK are in 4 – 6 interval [51]. At interaction with DNA TM ions provoke point defects such as double proton transfer in GC pairs, depurinization, create inter strand cross-links [27] which are the reason of intercalators ejection from DNA [51]. In its turn, the above influences the efficiency of nonradiative transfer of electronic excitation energy from donor (D) to acceptor (A).

The base of Förster mechanism of nonradiative transfer of electronic excitation energy is the so-called inductive-resonance transfer of energy from D to A, where dipole-dipole interaction dominates. According to Förster in our case the rate of energy transfer $k_{ET}$ is in direct ratio to donor emission quantum yield ($q_{oD}$), overlap integral donor emission spectra and extinction coefficient of acceptor molecule; and inversely to solvent refraction index in the fourth degree, distance between excited $D^*$ and A in the sixth degree, and $\tau_D$ is the donor emission decay time [52].

On the other hand, at inductive-resonance transfer of energy trough thin dielectric layers with alternative thickness d and permanent concentration of D and A the following correlation will be achieved [53, 54].

$$\frac{q_D}{q_{oD}} = 1 - \frac{q_A}{q_{oA}} = \left[1 + \left(\frac{d_0}{d}\right)^s\right]^{-1}, \qquad (7)$$

where s = 4, 6 and 2 are the energy transfers, consequently, for electric dipole – electric dipole, electric quadruple – electric dipole and magnetic dipole – electric dipole, $d_0$ – critical thickness of the layer. Correlation (1) in case of energy transfer from AO to EB intercalated in DNA double helix, can be re-written as follows:

$$\frac{q_{AO}}{q_{oAO}} = 1 - \frac{q_{EB}}{q_{oEB}} = \left[1 + \left(\frac{R_0}{R}\right)^s\right]^{-1} \quad (8)$$

where $q_{oD}$ is the quantum efficiency of donor fluorescence when the distance between energy donor and acceptor $R \to \infty$, $q_D$ at a given R, $q_{oA}$ is the quantum efficiency of sensibilized acceptor fluorescence at $R \to 0$ and $q_A$ at a given R The value $e_{ET} = 1 - q_{AO} / q_{oAO}$ was estimated as electron excitation energy transfer efficiency.

So, having definite concentration of intercalated pair D–A in DNA, and changing the concentration of DNA double helix the efficiency of energy transfer in the particular D-A pair can be sufficiently changed. Thus, the efficiency of energy transfer is proportional to DNA double helix site concentration. In Section 3 we show the validity of such approach for DNA-AO-EB ternary complexes.

Figure 15 shows fluorescence spectra of binary and ternary AO-DNA and AO-ED-DNA complexes where concentration of DNA changes the distance between donor AO and acceptor EB. AO and EB concentrations were constant and equal to 0.14 x $10^{-4}$ M. DNA concentration varied from 0.5 x $10^{-4}$ to 10 x $10^{4}$ M per base pair (*bp*). Figure 16 shows the ignition curves of AO fluorescence depending on the distance between AO and EB intercalated in DNA. The distance between donor and acceptor is given in nm and in *bp* units. The ignition curves of AO fluorescence are build in correspondence to Eq. (2) for s index equal to 4 , 6 or 2. An important characteristic of energy transfer process is Foerster distance $R_0$. At this distance, half of the donor molecules decay by energy transfer and the other half decays by the usual radiative and nonradiative rates [55].

$$R_0 = 0.211 \times 10^{-1}\left(\kappa^2 n^{-4} q_{oD}\, J(\lambda)\right)^{1/6} \text{ (in nm)} \quad (9)$$

This expression allows to calculate the Förster distance from the spectral properties ( $J(\lambda)$) of the donor and the acceptor and the donor quantum yield ($q_{oD}$), i.e. in terms of the experimentally known values taking into account the environment refractive index (n) and orientation factor ($\kappa^2$) of fluorescent molecules. In our case $J(\lambda) = 2.72 \times 10^{14}$ $M^{-1}(cm)^{-1}(nm)^4$, $q_{oD} = 0.75$ [56], $\kappa^2 = 2/3$. For the index of refraction we choose n = 1.6 [57]. Thus, the Förster distance is calculated as $R_0 = 3.5 \pm 0.3$ nm which is in good agreement with experimental value $R_0 = 3.9 \pm 0.3$ nm.

Table 3 presents calculated values for $R_0$ for different $n$ and $k^2$. As a refraction index some authors use $n$ = 1.33 (water) [58], others assumed $n$ to be 1.40, which is valid for biomolecules in aqueous solution [55] or $n$ = 1.60 in the case of DNA [57]. In all cases any dispersion effects are ignored. The term $k^2$ is a factor describing the relative orientation of donor and acceptor transition dipoles in space. $k^2$ is usually assumed to be equal to 2/3 which is appropriate for dynamic random average orientation of the donor and acceptor. When donor and acceptor are oriented in parallel planes, then $k^2 = 1/2$ (for details see [55]). Evidently the orientation factor $k^2$, and refractive index $n$ for different media do not affect significantly on $R_0$ value.

In Figure 16 we can see that theoretical curve describing electron excitation transfer for the case electric dipole – electric dipole (S = 4) corresponds to experimental data the best. Table 4 gives values for efficiency $e_{ET} = (1 - q_D/q_{oD})$ for AO intercalated in DNA depending on the distance between AO and EB given in *bp* units. The same table allows to estimate the distance between donor and acceptor in correspondence with the values of $e_{ET}$ evaluated from fluorescence spectra of FRET.

It is well known and it was shown in our work [27] that transition metal ions at interaction with DNA cause or participate in different conformational changes, e.g., Cu(II) ions initiate DNA transition from B to C form [59]; with the help of TM ions and ethanol B-Z transition can be initiated [27]. Effect of Ag(I) ions on DNA structure is known [41]. As far as in 1996 we discovered ejection of AO and EB intercalators from DNA. Besides, the rise of temperature in the solution causes melting of DNA double helix. In this connection it was interesting to investigate electron excitation transfer in D-A pairs intercalated in DNA under the effect of different stress factors with the aim of finding intact sites of DNA double helix.

## 5 Nanoscale donor-acceptor fluorescent probing of DNA double helix and stress factors: models and mechanisms.

The paragraph is devoted to the effects of Cu(II), Cu(I) and Ag(I) ions, silver nanoparticles, temperature and laser irradiation on electron excitation energy transfer, efficiency between intercalated in DNA fluorescence molecules of acridine orange (donor) and ethidium bromide (acceptor). Besides, the paragraph deals with the models and mechanisms of DNA formation.

## 5.1 Donor-acceptor energy transfer efficiency in DNA complexes with metal ions and silver nanoparticles

Figures 17, 18, 19 and 20 demonstrate the influence of Cu(II), Cu(I), Ag(I), AgNPs and laser irradiation ($\lambda$ =457 nm) on electron excitation energy transfer efficiency from AO to EB intercalated in DNA which is shown in the rise of effectiveness FRET. The estimated data are given in Table 4. There are also the data for the distance between AO and EB in *bp* units, as well as the relative concentrations of DNA sites applicable for intercalation. Figs. 17-20 also show that these ions quench AO fluorescence – the phenomena was studied by our group in [36,51] and it is connected with both dynamic quenching and the quenching caused by nonradiative transfer of excitation energy. On the one hand, Cu(II), Cu(I), Ag(I), Ag NPs and laser irradiation ($\lambda$ = 457 nm) quench AO fluorescence and on the other hand, increase FRET intensity. It obviously points out that there are different reasons for the phenomenon, in particular, FRET intensity is connected with the quality of double helix, i.e. stability constant of AO and EB with DNA.

## 5.2 Temperature effect on donor – acceptor energy transfer in DNA

Besides, temperature effect on FRET stability was investigated. Figure 21 shows heating effect on DNA solution located in a hermetic test tube in thermostat. Two ml of DNA solution was heated at various temperatures T = 50, 60, 70, 80 and 90$^o$C and then it was taken out and put into icy bath (T=0$^o$C). At 20$^o$C intercalator pair AO-EB was added to the solution and fluorescence spectra were registered.

Figure 22 shows fluorescence spectra for ternary complexes AO-DNA-EB. The complexes were prepared as follows: DNA solutions put in hermetic test-tubes were kept in vitro in thermal bath at the temperature 100$^0$ C for different periods of time (5, 10 and 20 min). Then they were quickly cooled in icy bath (T = 0$^o$C). After the procedure intercalator pair AO-EB was added to the solution and fluorescence spectra were registered. From the spectra given in Figures 21, 22 and with the application of Table 4 the effectiveness of FRET, the data for the distance between AO and EB in bp units, as well as the relative concentrations of DNA sites applicable for intercalation were estimated. The results are given in Table 5.

## 5.3 Laser irradiation and donor – acceptor energy transfer efficiency in DNA

Figures 23 and 24 show absorption spectra of AO molecules and binary complexes AO-DNA in water solutions before and after laser irradiation (λ = 457 nm). The given spectra distinctly present isosbestic points that show partial discoloration and destruction of AO from the one hand, and from the other hand, the transit of AO molecule from cation state to neutral one (N in meso-state: $N^{+}$-H→N of pyridine type). We should underline that the process of AO molecules intercalation in DNA 13 times decreases quantum yield of AO damages (discoloration and degradation) after 20 min laser irradiation ( λ =457nm).

Figures 25 and 26 show fluorescence spectra of AO molecules and binary complexes AO-DNA in water solutions before and after laser irradiation (λ = 457 nm). Analyzing the values of evaluated quantum yield of AO fluorescence in water solution we can draw a conclusion that the quantum yield value before irradiation 0.34 decreases to 0.1 after 20 min irradiation. In case when AO molecules are intercalated in DNA at the same irradiation conditions quantum yield of AO fluorescence is changed from 0.75 to 0.54.

Figure 27 presents changes in the values of AO molecule (λ = 502 nm) absorption spectra maximums in water solutions and in binary and ternary complexes with DNA- AO and DNA-AO-Cu(II), Cu(I), and Ag(I). Analyzing the data we can see that Cu (II) and Cu (I) ions decrease the destructive effect of irradiation while Ag(I) ions increase it.

Figure 28 shows the change of energy transfer efficiency from 13 base pairs to 7 base pair from AO to ethidium bromide intercalated in DNA after 20 min irradiation of binary DNA-AO complex (see Table 5).

## 5.4 New models and mechanisms of defect formation in DNA caused by stress factors

### 5.4.1 Double proton transfer in GC-pairs in DNA

The influence of $H^{+}$ ions on UV spectra of guanine and cytosine is known for a long time [60,61] as well as its explanation by keto-enolic and amino-imino transformations in guanine and cytosine, correspondingly. It should be noted that the mentioned transfers of proton ions between proton-donor and proton-acceptor groups N(1) and O(6) in guanine and N(4) and N(3) in cytosine are definitely performed with the help of the hydrated $H_2O$ molecules; they form a kind of cyclic structures with the above groups in guanine and cytosine. Because of self-congruent transition of protons from N(1) of guanine and N(4) of cytosine to molecule of hydrated $H_2O$, reorientation of $H_2O$ molecules take place in both cases.

Thus, we can state that the specific energy of H-bonds with two energy states and self-congruent transfer of protons in cyclic structures is a necessary and sufficient condition for keto-enol and amino-imino tautomeric transformations of polar organic molecules, particularly guanine, cytosine and adenine in solution. Water molecules act as a mediator in the process of forming the joint electron-proton complementary complex between the proton-acceptor/donor groups of the solute molecules and those of the solvent. As already mentioned, there is no need for water molecules as mediators in the DNA duplex because a pair of complementary bases always forms a cyclic structure.

Double proton transfer (DPT) in DNA is demonstrated spectroscopically as bathochromic shifts, weak hypochromic effects and small widening of the absorption band [27,36]. We shall discuss the mechanism in detail, using GC pairs as an example for the following reasons:

i. the effect of $H^+$ interactions with guanine and cytosine [41,62], as well as interaction between transition metal ions and guanine are easily observed in UV-spectra;
ii. spontaneous mutations of a genome occur more often at GC pairs than at AT pairs [63-65];
iii. guanine and cytosine are often populated with rare enol and imino forms [66-68];
iv. GC pairs are far less resistant to tunneling transitions compared to AT pairs [69-71].
v. in the DNA duplex the site of preferable binding of $H^+$ and metal ions is the endocyclic N7 of guanine located in the major groove [32,31].

In general, the keto-enol and amino-imine tautomeric transformations in GC pairs are conditioned by the electric charges on G-N1 and C-N3 of endocyclic nitrogens, which play the principal role in H-bonding of the base pairs. Figures 29 and 30 present electronic configurations of atoms participating in the formation of H-bonding in GC and AT pairs. It is interesting to consider here the disturbance of the electronic structure of guanine in the pair of bases in DNA duplex provoked by $H^+$ and transition metal ions.

When a positive ion interacts with a pyridine-type nitrogen (G-N7), a nonbonded pair of electrons located in a $sp^2$ hybrid orbital can significantly overlap the π-electronic system of the indole ring of guanine inducing a decrease of electronic density of the ring, including the endocyclic N1 nitrogen atom. This will lead to a decrease in the depth of potential energy from the N1 side of guanine and an increase of possible proton tunneling to N3 in cytosine, as illustrated in Figure 31.

### 5.4.2 Inter-strand cross-links in DNA double helix: ion sorption as a multistage adsorption process

Soft ions in particular $Cu^{+}$, $Ag^{+}$, $Pt^{++}$, $Hg^{++}$ ions are able to form the so-called inter-strand crosslinks in DNA [60]. Let's consider the process on the example of $Ag^{+}$. First, silver ions are adsorbed on DNA major groove (N7G or chelate complex N7G and O6G). At small silver ion concentration on DNA does not cause ejection of AO and EB [36]. On the other hand, silver ions at interaction with DNA induce double proton transfer in GC pair (see 5.4.1). Chelate complex with silver ions makes it easy to unfold DNA double helix with wrong Watson-Crick GC pair. Figure 29 presents electron configuration of GC atom pairs taking part in H-bonds before and after DPT. In the last process Guanine's atom O6 is in enol form, Nitrogen atom N1G is in pyridine state and N3 C in pyrole state. After unfolding of double helix in neutral water solution N3C atom cannot keep enol state in a long time and it should transfer into its usual pyridine state. At the same time silver ions can with definite possibility attack nitrogen atoms $N_1G$ still existing in pyridine state. During the following folding of double helix inter-cross link formation between N1G* and N3C takes place. This way the process of inter-cross link formation can be considered as such a simple process as: 1. Silver ion adsorption on DNA (N7G) and double proton transfer of GC pair with the life-time $\tau_1$, 2. Unfolding of double helix, formation of N1G* - $Ag^{+}$ binding, HN3C transfer to N3C and formation of link between $N_1G^*$-$Ag^+$ -$N_3C$. Total time of the process is $\tau_2$, 3. DNA folding with formation of stereoscopically distorted double helix with inter-cross links ($\tau_3$). So, in the case of DNA compound absorption process of inter-cross link formation can be reduced to a multi-stage adsorption process consisting of several simple adsorption processes named above with the total time of $\tau_1+\tau_2+\tau_3$. Scheme 1 and Figure 32 illustrate the said.

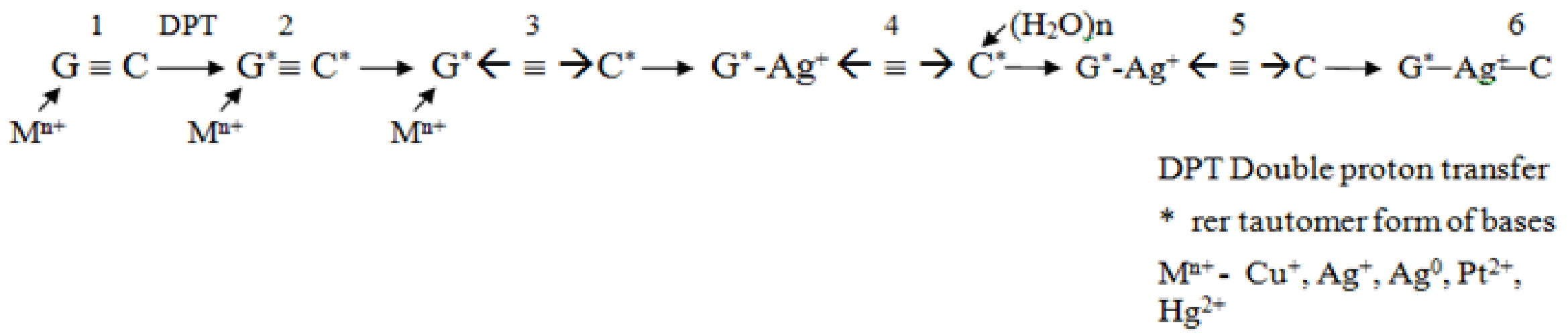

**Scheme 1.** The interstrand cross-link in DNA indused by silver ions.

### 5.4.3 Glycoside linkage (C(I) – N(9)) hydrolyze in DNA – depurinization

*Depurination.* Interaction of $H^+$ and $Me^{n+}$ with N(3) and N(7) of guanine in DNA with certain probability leads to DPT on the one hand and to hydrolysis of glycosidic linkage C(1)–N(9) on the other.

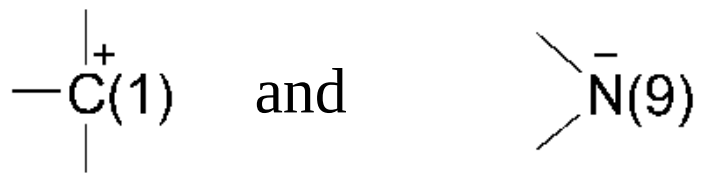

The reaction takes place with participation of dissociated molecules of $H_2O$ ($OH^-$ and $H^+$). Of course, $H^+$ approaches negatively charged $N^-$ as $OH^-$ goes to $C^+$:

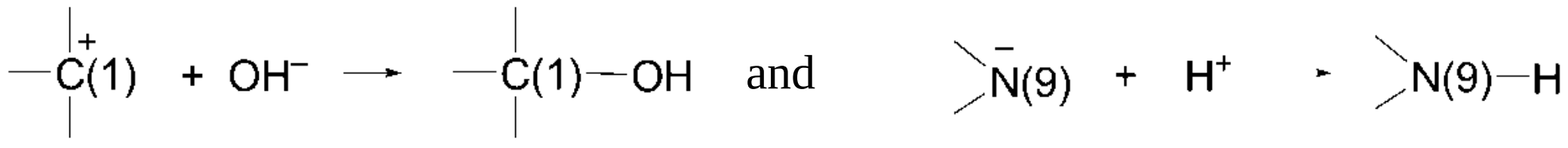

It needs to be mentioned that depurination can take place only in unwinded sites of DNA double helix. The dependence of the probability of unwinding of the central AU and GC base pairs in RNA double helix on the adjacent pairs was investigated in [22], and it was shown that the probability of opening of the G-C pair is minimum and equal to $0.3 \cdot 10^{-5}$ in sequence:

G–G–G<br>
• • •<br>
C–C–C

and the probability of opening of the A-U is maximum and equal to $120 \cdot 10^{-5}$ in sequence:

A–A–A<br>
• • •<br>
U–U–U

### 5.4.4 Phosphordiester linkage hydrolyze in DNA: photo-dynamic effect and Jablonski diagram

Hydrolyze reaction requires the presence of $H^+$ and $OH^-$ ions the concentration of which in water at pH 7 (natural pH for living cell) is $10^{-7}$M, i.e. the electric dissociation constant is $10^{-14}$. Naturally, at such concentrations the probability of hydrolyze is very small. *In vivo* phosphordiester link hydrolyze in DNA and RNA is carried out by nuclease ferments (DNase and RNase) and *in vitro* photoenergy dissipation in solution acts as a catalyst first of all in phosphordiester and glycoside links hydrolyze reactions Figure 33. It can not be excluded that in discoloration and degradation of aromatic compounds the hydrolyze reaction is also important. As life-time ($\tau$) of exited AO singlet state is ~$10^{-9}$ sec ($\tau_{A0} = 15 \cdot 10^{-9}$, $\tau_{A0\text{-}DNA} = 5.2 \cdot 10^{-9}$ sec [72] ) the possibility that oxygen molecule ($[O_2]=1.3 \cdot 10^{-6}$ M at t=25 $^0$C, see table 2) can collide AO molecule is too small. At the same time triplet excited state is unusual for AO molecule at room temperature [73,74].

So in DNase at photodynamic effect in solutions the principal oxidant is not an oxygen molecule but $H^+$ ions. It is also connected to the fact, which we have shown, that for $H^+$ ions mobile-adsorption state of DNA is typical. It should be noted that at photodynamic effect $H^+$ local concentration depends on electrolytic dissociation of water molecule. So at photon absorption by chromophore a part of energy dissipates into heat as a result of conversion and it should increase the constant of electrolytic dissociation (see Figure 34). For instance, $100^0$ temperature raise causes increase of water dissociation constant by 3 orders (Arrhenius equation). It's interesting how many $H^+$ and $OH^-$ ions can one photon ($\lambda$ = 457 nm) generate if it totally dissipates into heat. As electrolytic dissociation reaction activation energy requires 19.5 kcal/M the energy of absorbed photon is enough for 3 water molecules to undergo electrolytic dissociation. In the place where AO exited molecule undergoes total energy dissipation we can get high local concentration of $H^+$ and $OH^-$ ions in immediate proximity from DNA, and thus hydrolyze possibility of both phosphordiester and glycoside is significant. In particular, phosphor-di-ester links a) P-$O_{(3')}$ and b) $O_{(5')}$-P can be presented as two schemes.

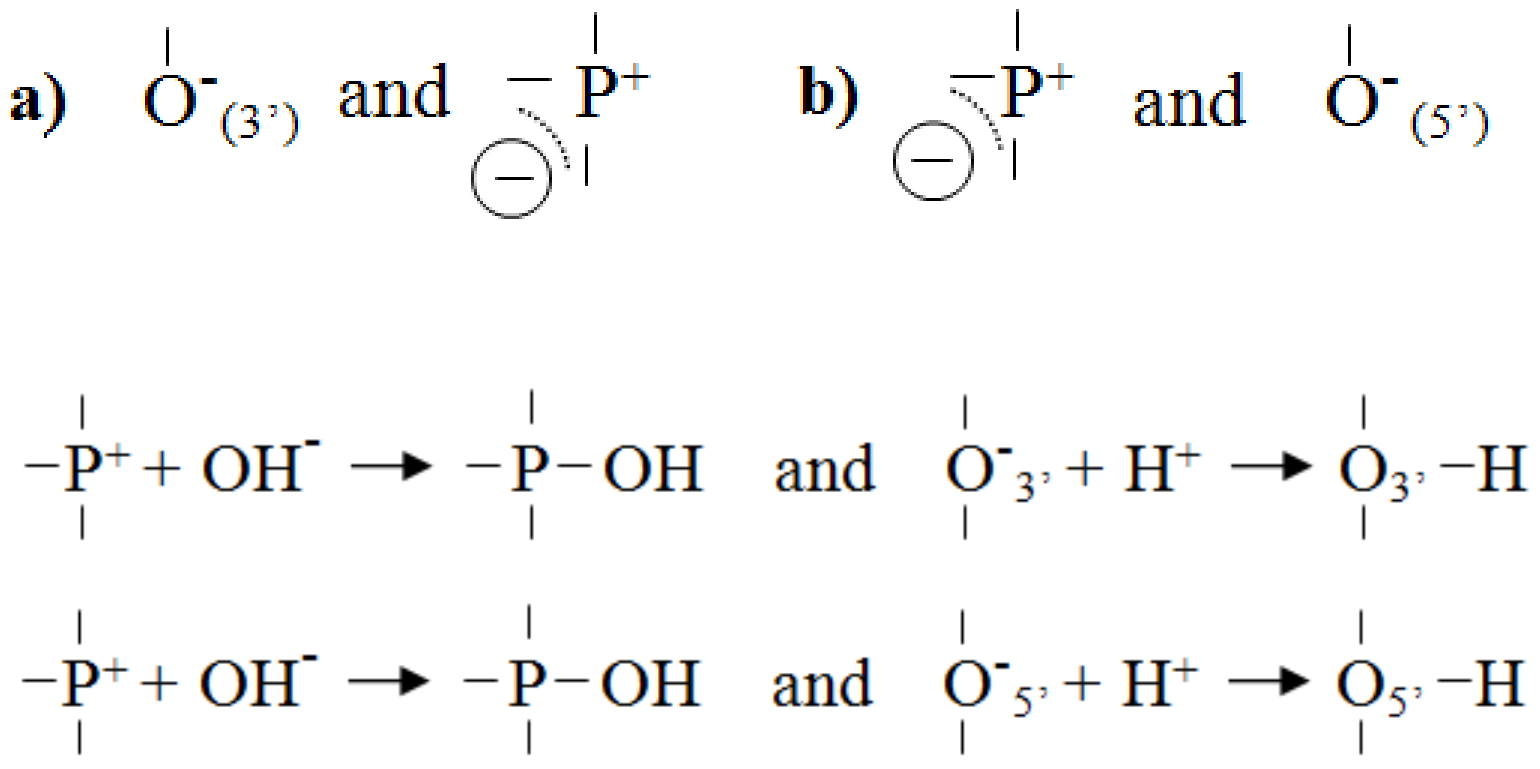

**6 Conclusions.**

Using spectrophotometry and thermodynamic approaches we have shown that 1) at interaction with DNA, silver nano particles with the size 1-2 nm (AgNPs) are adsorbed on it and only partial corrosion of nanoparticles at the level of $Ag^+$ ions is observed; 2) AA reduces silver ions in ternary complex DNA- $AgNO_3$- AgNPs. Thus, AgNPs activate the process of quick reduction of $Ag^+$ ions to silver atoms; 3) at photo-irradiation ($\lambda$ = 436 nm or full spectrum of visible band) desorption of silver atoms from the surface of AgNPs takes place. The atoms are first adsorbed on the surface of DNA and then penetrate inside the double helix (cross links between complementary DNA base pairs) making prolate stretched structure (AgNPs absorption spectrum width is changed from 140 nm to 360 nm at half-height); 4) kinetic study of photo-desorption makes it possible to determine desorption rate constant $k_d$ and adsorption heat $Q_a$ that are equal to $k_d \cong 9 \times 10^{-5}$ $s^{-1}$; $Q_a \geq 80$ kJ/M $Ag^0$ for AgNPs bound with DNA; 5) AgNPs represent liquid drops which moisture the DNA surface at interaction. At photo-irradiation of AgNPs – DNA complex DNA dependant conformational transition takes place due to fast and intensive heating.

The nano-scale method of laser induced fluorescence resonance energy transfer (FRET) to donor-acceptor intercalator pair for quantitative and qualitative study of stability quality DNA double helix in solution in real time is offered. The approach is based on the example of acridine orange molecule (donor) and ethidium bromide (acceptor) intercalated in DNA.

It is shown that ions Cu(II), Cu(I), Ag(I) and Ag NPs, laser irradiation of AO and the effect of heating decrease the concentration of undamaged areas of DNA double helix, i.e. the sites able to intercalate dye molecules such as AO and EB.

FRET radii were experimentally estimated in background electrolyte solution (0.01 M $NaNO_3$) and proved to be 3.9 ± 0.3 nm and the data are in satisfactory agreement with the theoretically calculated value $R_0 = 3.5 \pm 0.3$ nm.

FRET method allows to estimate the concentration of double helix areas with high quality stability applicable for intercalation in DNA after it was subjected to stress effect. It gives the opportunity to compare various types of DNAs with 1) different origin; 2) various damage degrees; 3) being in various functional state.

Alternative model and mechanisms of photodynamic effect on DNA in solutions are proposed. They are based on photo energy degradation in solutions. The energy activates electrolytic dissociation of water molecules on $H_3O^+$ and $OH^-$ and acts as a catalyst for hydrolyse reactions of phosphor-di-ester and glycoside linkages.

## 7 Experimental Part.

### *7.1 Materials*

7.1.1. DNA. In our tests we used the Calf thymus DNA (40% GC), 'Sigma'. The concentration of nucleic acids was determined by UV absorption using molar extinction coefficients ($\varepsilon$ = 6600 $cm^{-1}$ $M^{-1}$ at $\lambda$ = 260 nm). The double helix structure of the polymers was proved by their hyperchromicity (> 30%) and their typical thermal denaturation transition. (measured in 0.01 M $NaNO_3$, pH ≅ 6.0). pH was checked by pH meter HANNA Instruments pH213.

7.1.2. Intercalators. Acridine orange (AO) was purchased from 'Sigma'. The concentration of the dye was determined colorimetrically at the isobestic point of the monomer-dimer system (λ=470 nm) using the molar extinction coefficients (ε = 43 300 $cm^{-1}M^{-1}$). Ethidium bromide (EB) was also purchased from 'Sigma'. The concentration of the dye was determined colorimetrically (ε = 5600 $cm^{-1}M^{-1}$ at λ = 480 nm).

7.1.3. Ions. We used chemically pure copper chloride. Bidistillate water served as a solvent. In tests with $Ag^+$ ions chemically pure salts $AgNO_3$ were used and $NaNO_3$ served as background electrolytes.

7.1.4. Nanoparticles. Colloidal silver suspension with particle sizes of 1-2 nm in distilled water (200μg/ml) was purchased from DDS Inc., D/B/A, Amino Acid & Botanical Supply P.O Box 356, Cedar Knolls, NJ 07927.

7.2. *Instrumentation*

7.2.1. Absorption spectra of DNA complexes with intercalators AO and EB were registered in real time using CCD spectrometer AvaSpec ULS 2048-USB2. It should be underlined that registration of fluorescence spectra excited by laser irradiation is necessary to carry out in real time as at such excitation of intercalators, AO in particular, its fast photo-oxidation takes place

7.2.2. Diode laser SDL–475–100T (Shanghai Dream Lasers Technology Co., Ltd.) was used for irradiation and excitation (λ = 457 nm with optical beam cross-section 2 mm, and P=200 mW) of laser induced fluorescence spectra.

7.2.3 Photoirradiation was carried out in reactor with the fixed light beam in 1cm rectangular fluorescent quartz cell. In the same cell with the interval of 5 min absorption spectra of irradiated solutions were registered by AvaSpec spectrometer. Before each absorption registration the cell was shut to protect the solution from photo irradiation. Registration time was 8 msec. As a source of radiation xenon arch discharge lamp with rating of 35 W in glass balloon was used. To irradiate the solution water filter and light filter with light wave transmission $\lambda$ = 436 nm were used. Radiation power in the cell was 300 mW for thermal water filter and 15 mW for water filter matched with light filter ($\lambda$ = 436 nm).

**Acknowledgements**

The authors express their gratitude to Prof. *Henri Kagan*, Paris-Sud University, Orsay, for his permanent kind attention to our work. They also express gratitude to Prof. *Jamlet Monaselidze* for usefull discussions, and to Mrs. *G. Nijaradze* for her help in preparing the manuscript.

The work was partly supported by the grants of Shota Rustaveli National Science Foundation: GNSF/ST09_508_2-230, GNSF41/14, GNSF 12/24, GNSF-STCU #5622.

This work is supported by the IPP/ISTC A-1951 Project.

## Figure captions

**Figure 1.** Fluorescence spectra of ternary complex DNA-AO-EB.[AO]-$3\cdot10^{-5}$ M, [EB]-0 M (1); [AO]-$2\cdot10^{-5}$ M , [EB]-$1.0\cdot10^{-5}$ M (2); [AO]-$1.5\cdot10^{-5}$ M , [EB]-1.5 $10^{-5}$ M (3); [AO]-$1\cdot10^{-5}$ M, [EB]=$2\cdot10^{-5}$ M (4). [DNA]-$10^{-4}$ M (P), [NaCl]-0.01 M, pH7.

**Figure 2.** Absorption spectra of AgNPs and AgNPs-DNA complexes. [AgNPs] – $0.72\cdot10^{-4}$ M ($Ag^0$), [DNA] –$1.6\cdot10^{-4}$ M (P), [$NaNO_3$] –$10^{-2}$ M.

**Figure 3.** First derivative of absorbtion spectra of AgNPs and AgNPs-DNA. The blue shift is evident.

**Figure 4.** Absorption spectra of: AgNPs after various ion strength application. [AgNPs]-$0.7\cdot10^{-4}$M ($Ag^0$), [$NaNO_3$]-$10^{-2}$M ($C_1$); [$NaNO_3$]-$3\cdot10^{-1}$M ($C_2$).

**Figure 5.** Absorption spectra of: AgNPs and DNA after various ion strength application. [AgNPs]-$0.7\cdot10^{-4}$M ($Ag^0$), [DNA]-$1.8\cdot10^{-4}$ M(P), [$NaNO_3$]-$10^{-2}$M ($C_1$); [$NaNO_3$]-$3\cdot10^{-1}$M ($C_2$).

**Figure 6.** Absorption spectra of AgNPs and the following complexes DNA-$AgNO_3$-AgNPs and DNA-$AgNO_3$-AgNPs-AA. [AgNPs]-$0.7\cdot10^{-4}$M ($Ag^0$), [DNA]-$1.8\cdot10^{-4}$ M(P), [$AgNO_3$]-$0.7\cdot10^{-4}$M, [AA]-$1.4\cdot10^{-4}$M, [$NaNO_3$]-$10^{-2}$M.

**Figure 7.** Spectra for silver ion reduction in complexes DNA-$AgNO_3$-AA. [DNA]-$5.9\cdot10^{-4}$ M(P), [$AgNO_3$]-$1.2\cdot10^{-4}$M, [AA]-$2.4\cdot10^{-4}$M, [$NaNO_3$]-$10^{-2}$M.

**Figure 8**. Absorption spectra of AgNPs before (1) and after (2) heating (15 min incubation at T=373 K) [AgNPs]–$0.72\cdot10^{-4}$ M ($Ag^0$), [$NaNO_3$]–$10^{-2}$ M.

**Figure 9**. Absorption spectra of AgNPs before and after irradiation (5min interval). [AgNPs] – $1.94\cdot10^{-4}$ M ($Ag^0$), [$NaNO_3$] – $10^{-2}$ M.

**Figure 10.** Absorption spectra of AgNPs-DNA before and after irradiation (5min interval). [AgNPs] – $1.94\cdot10^{-4}$ M ($Ag^0$), [DNA] – $1.6\cdot10^{-4}$ M (P), [$NaNO_3$] – $10^{-2}$ M.

**Figure 11.** Kinetic curve of photo-desorption in $M_t/M_e$ and $t^{1/2}$ coordinates for AgNPs bound with DNA.

**Figure 12.** Kinetic curve of photo-desorption in ln [$M_e/(M_e - M_t)$] and $t$ coordinates for AgNPs bound with DNA.

**Figure 13.** Model of DNA moisturizing by silver nanoparticles.

**Figure 14.** **(a)** Structure of acridine orange and **(b)** Structure of ethidium bromide.

**Figure 15.** Fluorescence spectra of double and ternary DNA-AO and DNA-AO-EB complexes where concentration of DNA changes the distance between donor AO and acceptor EB. (a) DNA-AO [DNA]-$2.8\cdot10^{-4}$ M (P), [AO]-$0.14\cdot10^{-4}$, [$NaNO_3$]-$10^{-2}$ M. (b) DNA-AO-EB [DNA]-$10\cdot10^{-4}$ M (P), [EB]-$0.14\cdot10^{-4}$ M, (c) DNA-AO-EB [DNA]-$8\cdot10^{-4}$; (d) DNA-AO-EB [DNA]-

$6\cdot10^{-4}$; (e) DNA-AO-EB [DNA]-$4\cdot10^{-4}$; (f) DNA-AO-EB [DNA]-$2.8\cdot10^{-4}$; (g) DNA-AO-EB [DNA]-$1.4\cdot10^{-4}$; (h) DNA-AO-EB [DNA]-$10^{-4}$.

**Figure 16.** Ignition curves of AO fluorescence depending on the distance between AO and EB intercalated in DNA

**Figure 17.** Influence of Cu(II) and laser irradiation ($\lambda$ = 457 nm) on electron excitation energy transfer effectiveness from AO to EB intercalated in DNA. (a) DNA-AO; (b) DNA-AO-EB; (c) DNA-AO-Cu(II); (d) DNA-AO-Cu(II) 10 min irradiation; (e) DNA-AO-Cu(II)-EB; (f) DNA-AO-Cu(II) 10 min irradiation +EB. [DNA]-$9.6\cdot10^{-4}$ M (P), [AO]-$0.14\cdot10^{-4}$M, [EB]-$0.14\cdot10^{-4}$ M, [$CuCl_2$]$0.14\cdot10^{-4}$ M, [$NaNO_3$]-$10^{-2}$ M.

**Figure 18.** Influence of Cu(I) ([ascorbic acid]/[$Cu^{2+}$] 2:1) and laser irradiation ($\lambda$ = 457 nm) on electron excitation energy transfer effectiveness from AO to EB intercalated in DNA. (a) DNA-AO; (b) DNA-AO-EB; (c) DNA-AO-Cu(I); (d) DNA-AO-Cu(I) 10 min irradiation; (e) DNA-AO-Cu(I)-EB; (f) DNA-AO-Cu(II) 10 min irradiation +EB. [DNA]-$9.6\cdot10^{-4}$ M (P), [AO]-$0.14\cdot10^{-4}$, [EB]-$0.14\cdot10^{-4}$ M, [$CuCl_2$]-$0.14\cdot10^{-4}$ M, [AA]-$0.24\cdot10^{-4}$ M, [$NaNO_3$]-$10^{-2}$ M.

**Figure 19.** Quenching of fluorescence by $Ag^+$ ion in DNA-AO-EB complex. (a) DNA-AO, (b) DNA-AO-EB- $Ag^+$ ($C_1$), (c) DNA-AO-EB-$Ag^+$ ($C_2$), (d) DNA-AO-EB-$Ag^+$ ($C_3$). [DNA]-$2.8\cdot10^{-4}$ M (P), [AO]-$0.14\cdot10^{-4}$ M, [EB]-$0.14\cdot10^{-4}$ M, [$Ag^+$]-0 ($C_1$), [$Ag^+$]-$6.0\cdot10^{-6}$ M ($C_2$), [$Ag^+$]-$30.0\cdot10^{-6}$ M ($C_3$), [$NaNO_3$]-$10^{-2}$ M. $\lambda$=460 nm

**Figure 20** Quenching of fluorescence by AgNPs in DNA-AO-EB complex. (a) DNA-AO, (b) DNA-AO-EB-AgNPs ($C_1$), (c) DNA-AO-EB-AgNPs ($C_2$), (d) DNA-AO-EB-AgNPs ($C_3$). [DNA]-$2.8\cdot10^{-4}$ M (P), [AO]-$0.14\cdot10^{-4}$ M, [EB]-$0.14\cdot10^{-4}$ M, [AgNPs]-0 ($C_1$), [AgNPs]-$6.0\cdot10^{-6}$ M ($C_2$), [AgNPs]-$18.0\cdot10^{-6}$ M ($C_3$), [$NaNO_3$]-$10^{-2}$M. $\lambda$=460 nm.

**Figure 21.** Heating effect on DNA solution located in a hermetic test-tube in thermostat at various temperatures t= 50, 60, 70, 80 and 90 $^o$ C. (a) DNA-AO, (b) DNA-AO-EB, (c) DNA-AO-EB 50$^o$C, (d) DNA-AO-EB 60$^o$C, (e) DNA-AO-EB 70$^o$C, (f) DNA-AO-EB 80$^o$C, (g) DNA-AO-EB 90$^o$C. [DNA]-$2.8\cdot10^{-4}$ M (P), [AO]-$0.14\cdot10^{-4}$ M, [EB]-$0.14\cdot10^{-4}$ M, [$NaNO_3$]-$10^{-2}$ M.

**Figure 22.** Fluorescence spectra for ternary complexes AO-DNA-EB. The complexes were prepared as follows: DNA solutions put in hermetic test-tubes were kept in vitro in thermal bath at the temperature $100^0$ C for different periods of time (5, 10 and 20 min). (a) DNA-AO, (b) DNA-AO-EB (0 min), (c) DNA-AO-EB (5 min), (d) DNA-AO-EB (10min), (e) DNA-AO-EB (20min). [DNA]-$2.8\cdot10^{-4}$ M (P), [AO]-$0.14\cdot10^{-4}$ M, [EB]-$0.14\cdot10^{-4}$ M, [$NaNO_3$]-$10^{-2}$ M.

**Figures 23** Absorption spectra of AO molecules in water solutions before and after laser irradiation ($\lambda$ = 457nm) are shown. **[**AO]-$0.7\cdot10^{-5}$ M, [$NaNO_3$]-$10^{-2}$ M.

**Figures 24.** Absorption spectra for binary complexes AO-DNA in water solutions before and after laser irradiation (λ = 457nm) are shown. [DNA]-$7\cdot10^{-4}$ M (P), [AO]-$0.14\cdot10^{-4}$ M, [$NaNO_3$]-$10^{-2}$ M.

**Figures 25.** Fluorescence spectra of AO molecules in water solutions before and after laser irradiation (λ = 457 nm). [AO]-$0.7\cdot10^{-5}$ M, [$NaNO_3$]-$10^{-2}$ M.

**Figures 26.** Fluorescence spectra for binary complexes AO-DNA in water solutions before and after laser irradiation (λ = 457 nm). [DNA]-$7\cdot10^{-4}$ M (P), [AO]-$0.14\cdot10^{-4}$ M, [$NaNO_3$]-$10^{-2}$ M.

**Figure 27.** Laser irradiation effect on AO molecules in binary and ternary complexes AO-DNA, AO-DNA-Cu(II), AO- DNA-Cu(I), DNA-AO-Ag and AO. ▶— AO-DNA-Cu(I); ●— AO-DNA-Cu(II), ▷ — AO-DNA, ■— DNA-AO-Ag(I) □—AO. [DNA]–$0.7\cdot10^{-3}$ M (P), [AO]–$0.7\cdot10^{-4}$ M, [EB]–$0.7\cdot10^{-4}$ M, [AA]–$1.4\cdot10^{-4}$ M, [$Ag^+$]–$0.7\cdot10^{-4}$ M, [$CuCl_2$]–$0.7\cdot10^{-4}$ M, [$NaNO_3$]–$10^{-2}$ M.

**Figure 28.** Influence of laser irradiation (λ = 457 nm) on electron excitation energy transfer effectiveness from AO to EB intercalated in DNA. (a) AO- DNA; (b) AO- DNA-EB; (c) AO-DNA irradiation 25 min; (d) AO-DNA irradiation 25 min + EB. [DNA]-$7\cdot10^{-4}$ M (P), [AO]-$0.14\cdot10^{-4}$ M, [EB]-$0.14\cdot10^{-4}$ M, [$NaNO_3$]-$10^{-2}$ M.

**Figure 29.** The electron configuration of atoms of G-C piars taking part in H-bonds before and after DPT.

**Figure 30.** Double proton transfer (DPT) in a G-C part of DNA.

**Figure 31.** Hypothetical function of potential energy of N-H···N type H-bonds.

**Figure 32.** The interstrend cross-link in DNA indused by silver ions. 1. Regular Wotson-Crick pair, 2. Wrong Wotson-Crick pair, 3. Open pair, 4. Silver ions start transfer together with water molecules in GC open wrong pair from N7 to N1, 5. In the open pair cyclic water molecules system makes proton transfer from N3 to N4 in cytosine, 6**. The interstrend cross-link.**

**Figure 33.** Segment of polynucleotide DNA change 1- glycoside link, 2- and 3- phosphordiester links.

**Figure 34.** Scheme of electron-vibration-ratation levels of intramolecule transitions in compound organic molecules and siglet-singlet energy transfer between donor and acceptor.

**Tables:**

**Table 1.** Characteristics of Inner Movements in DNA

| Type of movement | Time, *sec* | Excitation energy, *kcal/mol* |
|---|---|---|
| 1. Various small – amplitude oscillating movements of atoms (about 0,1 $\mathring{A}$) inside the components of DNA | $10^{-14}$—$10^{-13}$ ($3500$-$300 cm^{-1}$) | RT(T=$300^{o}$K)= 0.6(210 $cm^{-1}$) |
| 2. Limited movements of phosphates, sugars and nucleobases relative to the equilibrium position, torsional and flexural oscillations of the double helix | $10^{-10}$—$10^{-8}$ | |
| 3. Large –amplitude movements of phosphates, sugars and bases occurring in connection with the transition of the double helix from one form to another | $10^{-7}$—$10^{-5}$ | 5-6 for $B \rightarrow A$ transitions and<br>21 for $B \rightarrow Z$ transitions [21] |
| 4. Change of free energy $\Delta G$ needed for opening of central pairs in double chain RNA at $25^{o}$C, kcal/mol bases [22]<br>G C ( G —G —G / \| \| \| / C – C – C ) and<br>AU ( A —A —A / \| \| \| / U – U – U ) | $\sim 3 \cdot 10^{-6}$ *)<br>$\sim 10^{-8}$ *) | 7.5<br>4.0 |

*) The time of pair opening is evaluated by formulae (3)

**Table 2** Velues for reaction rate constants of hydrated with $Ag^+$, $H_3O^+$, $NO_3^-$ ions and $O_2$, $H_2O$ and DNA molecules

| *Ions and molecules* | *pH* | $k_{e/aq}$ $(M^{-1}s^{-1})$ | $\Delta Ga^{-1}$ | *Concentration in solution (M)* |
|---|---|---|---|---|
| $Ag^+$ | 7 | $3.2 \cdot 10^{10}$ | 1.468 | $1.1 \cdot 10^{-4}$ |
| AgNPs | 6 | - | - | $0.74 \cdot 10^{-4}$ ($Ag^0$) |
| $H_3O^+$ | 4-5 | $2.36 \cdot 10^{10}$ | 1.157 | $0.3 \cdot 10^{-4}$ |
| $NO_3^-$ | 7 | $1.1 \cdot 10^{10}$ | 0.757 | $10^{-2}$ |
| $H_20$ | 8.4 | $1.6 \cdot 10^1$ | 0.074 | |
| $Na^+$ | 7 | $<10^5$ | 0.121 | $0.8 \cdot 10^{-4}$ |
| $O_2$ | 7 | $1.88 \cdot 10^{10}$ | 1 | $1.33 \cdot 10^{-6}$ |
| DNA | 7 | $\sim 10^8$ | 0.242 | $2 \cdot 10^{-4}$ (P) |

**Table 3.** Calculated values of $R_0$ for different *n* and $k^2$

| N | $R_0(bp)$ | | $R_0(nm)$ | |
|---|---|---|---|---|
| | $\kappa^2 = \frac{2}{3}$ | $\kappa^2 = \frac{1}{2}$ | $\kappa^2 = \frac{2}{3}$ | $\kappa^2 = \frac{1}{2}$ |
| 1.33 | 11.72 | 11.17 | 3.98 | 3.79 |
| 1.40 | 11.32 | 10.79 | 3.85 | 3.67 |
| 1.60 | 10.36 | 9.87 | 3.52 | 3.35 |

**Table 4.** Efficiency of energy transfer $e_{ET} = (1 - q_D/q_{oD})$ from AO donor to EB acceptor depending on the distance R between them.

| $\mathbf{e_{ET}}$ | **R (*bp*)** | $\mathbf{e_{ET}}$ | **R (*bp*)** |
|---|---|---|---|
| 0.999 | 1 | 0.164 | 16 |
| 0.995 | 2 | 0.135 | 17 |
| 0.985 | 3 | 0.112 | 18 |
| 0.963 | 4 | 0.093 | 19 |
| 0.927 | 5 | 0.078 | 20 |
| 0.872 | 6 | 0.066 | 21 |
| 0.800 | 7 | 0.055 | 22 |
| 0.715 | 8 | 0.047 | 23 |
| 0.622 | 9 | 0.040 | 24 |
| 0.529 | 10 | 0.035 | 25 |
| 0.442 | 11 | 0.030 | 26 |
| 0.365 | 12 | 0.026 | 27 |
| 0.300 | 13 | 0.023 | 28 |
| 0.245 | 14 | 0.020 | 29 |
| 0.201 | 15 | 0.018 | 30 |

**Table 5.** Cu(II), Cu(I), Ag(I) ions, AgNP, laser irradiation (λ= 457 nm) and heating effects on $e_{ET}$[1)] and $C_{dh}^{st}/C_0 = R_{AO-EB}^{st}/R_{AO-EB}^{0}$[2)]

| **Stress factor for DNA - AO - EB** [3)] | $\mathbf{e_{ET}}$ (%) | $\mathbf{R_{AO-EB}}$ (***bp***) [4)] | $\frac{\mathbf{C_{dh}^{st}}}{\mathbf{C_0}} = \frac{\mathbf{R_{AO-EB}^{st}}}{\mathbf{R_{AO-EB}^{0}}}$ |
|---|---|---|---|
| - | 20 | 15 | 1 |
| Cu(II) | 58 | 10-9 | 0.63 |
| Cu(I) | 53 | 10 | 0.67 |
| Ag(I) | 67 | 8-9 | 0.57 |
| AgNPs (1) | 62 | 9 | 0.6 |
| AgNPs (2) | 76 | 8-7 | 0.5 |
| **Heating** | | | |
| $20^0$C | 20 | 15 | 1 |
| $50^0$C | 48 | 10-11 | 0.7 |
| $60^0$C | 52 | 10 | 0.67 |
| $70^0$C | 76 | 8-7 | 0.5 |
| $80^0$C | 81 | 7 | 0.47 |
| $90^0$C | 82 | 7 | 0.47 |
| **Boiling** | | | |
| $20^0$C | 30 | 13 | 1 |
| $100^0$C, 5 min | 80 | 7 | 0.54 |
| $100^0$C, 10 min | 87 | 6 | 0.46 |
| $100^0$C, 20 min | 95 | 4 | 0.31 |
| **laser irradiation** | | | |
| DNA-AO 25 min | 30 | 13 | 0.54 |
| Cu(II) (10 min*)* | 77 | 8-7 | 0.5 |
| Cu(I) (10 min*)* | 81 | 7 | 0.47 |

[1)]Foerster resonance electron excitation energy transfer from donor AO to acceptor EB. [2)] relative concentration of DNA double helix areas applicable for AO and EB intercalation, where $C_{dh}^{st}$ is concentration of double helix areas in *bp* left after stress effect, $C_0$ – initial DNA concentration in M *bp*, $R_{AO-EB}^{0}$ – distance between AO and EB at initial DNA concentrations, $R_{AO-EB}^{st}$ – distance between AO and EB after stress; [3)] different effects on DNA double helix; [4)] $R_{AO-EB}$ – distance between AO and EB in *bp* units evaluated from efficiency $e_{ET}$ ( see Table 4).

Figure 01

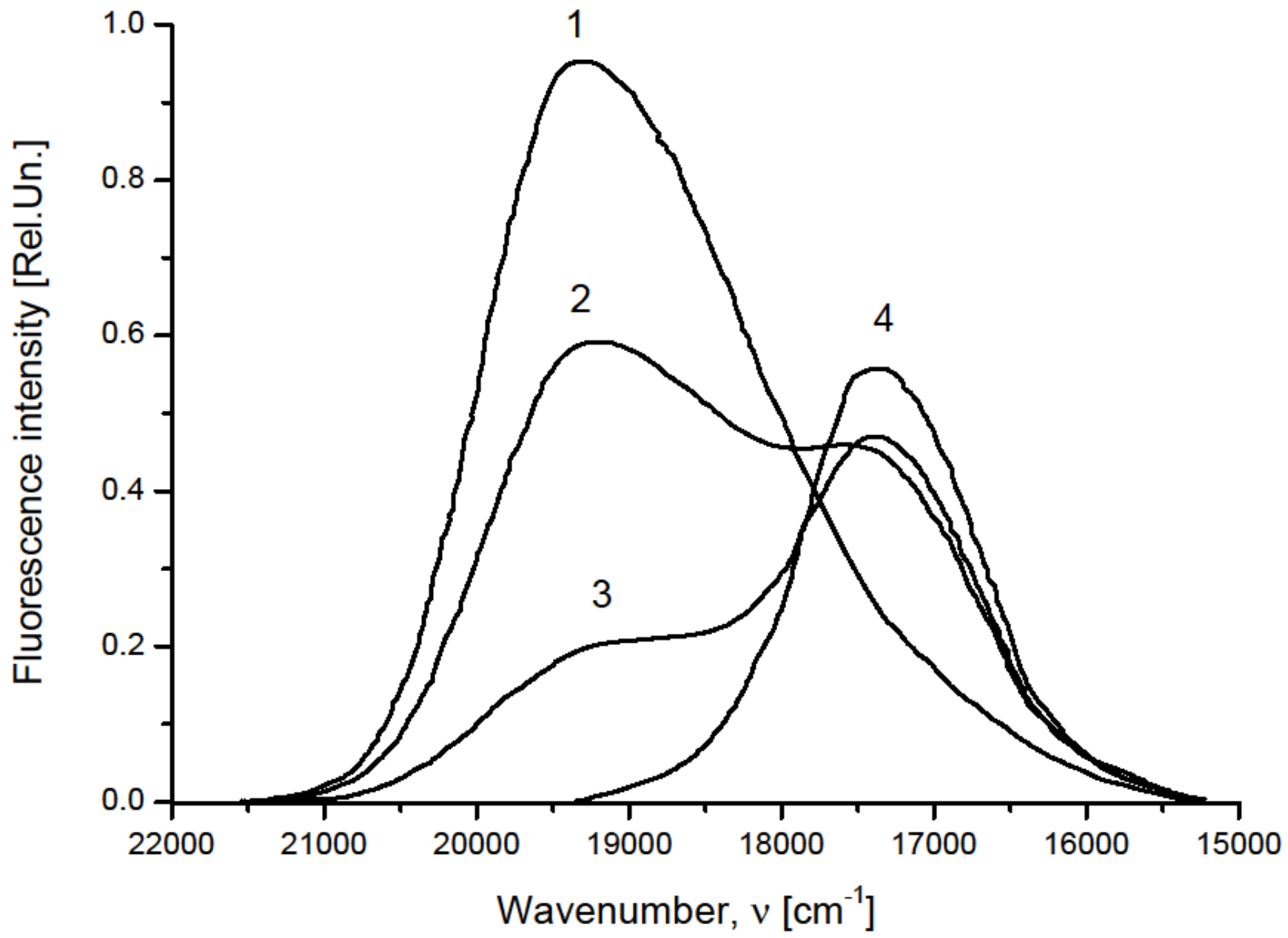

Figure 02

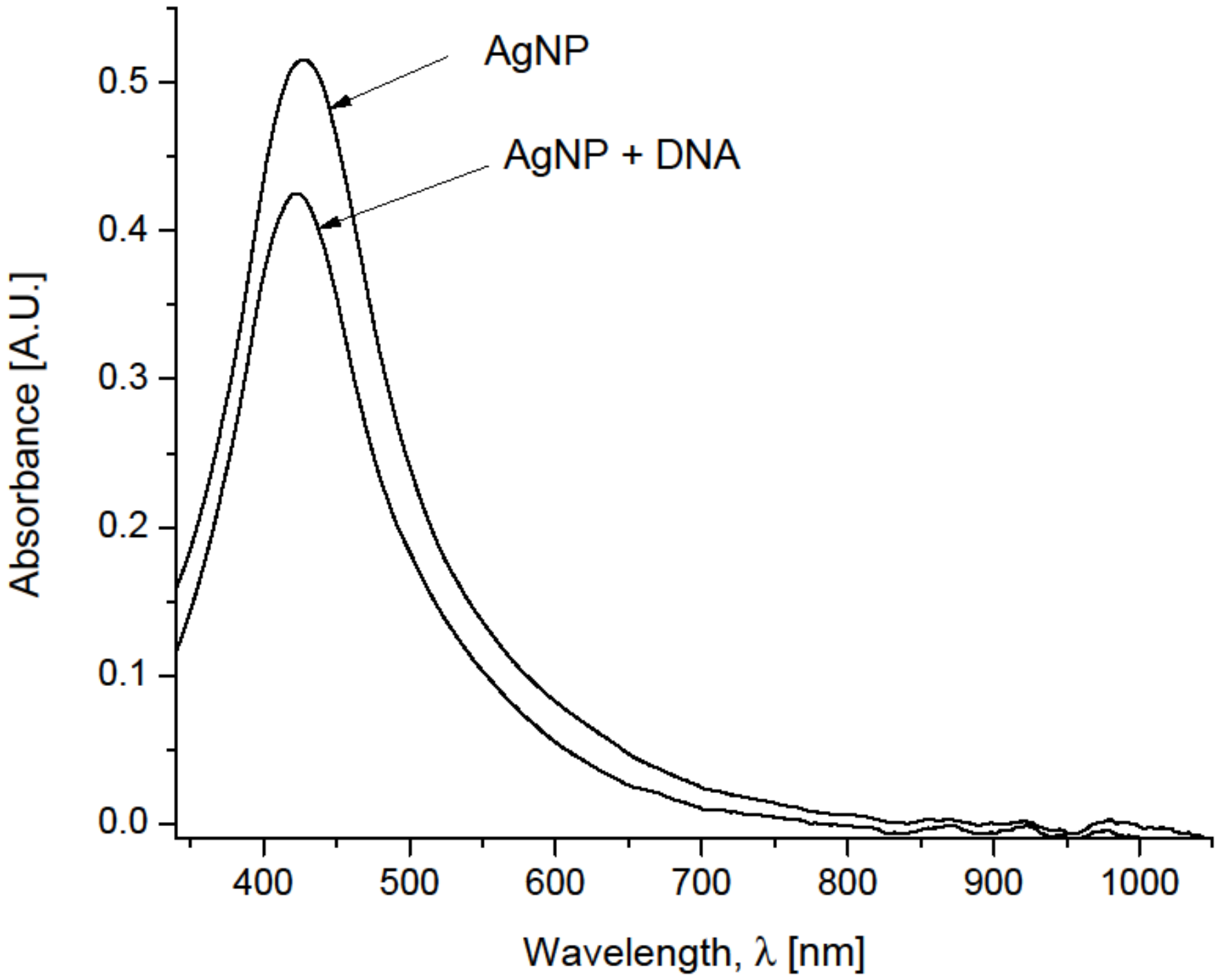

Figure03

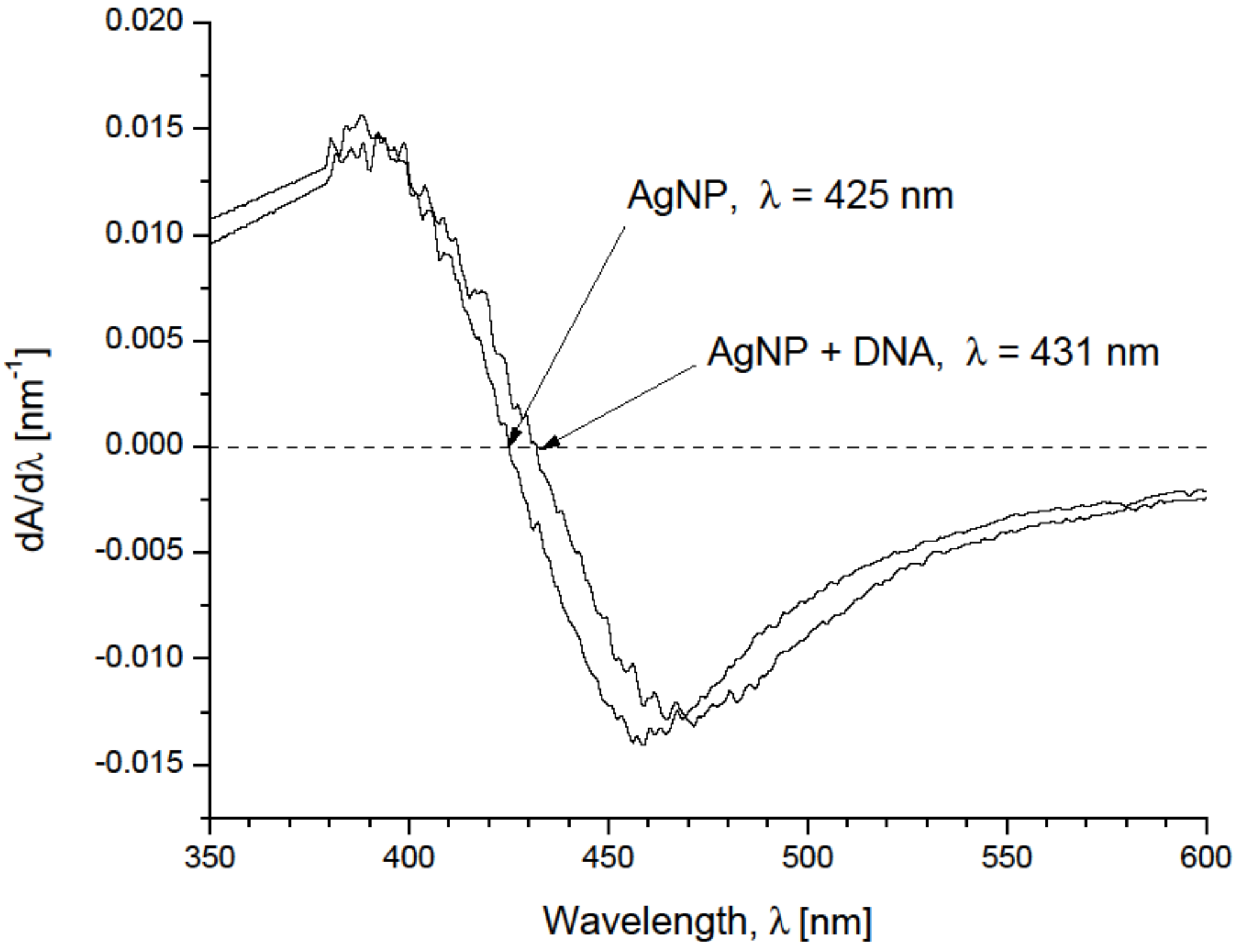

Figure04

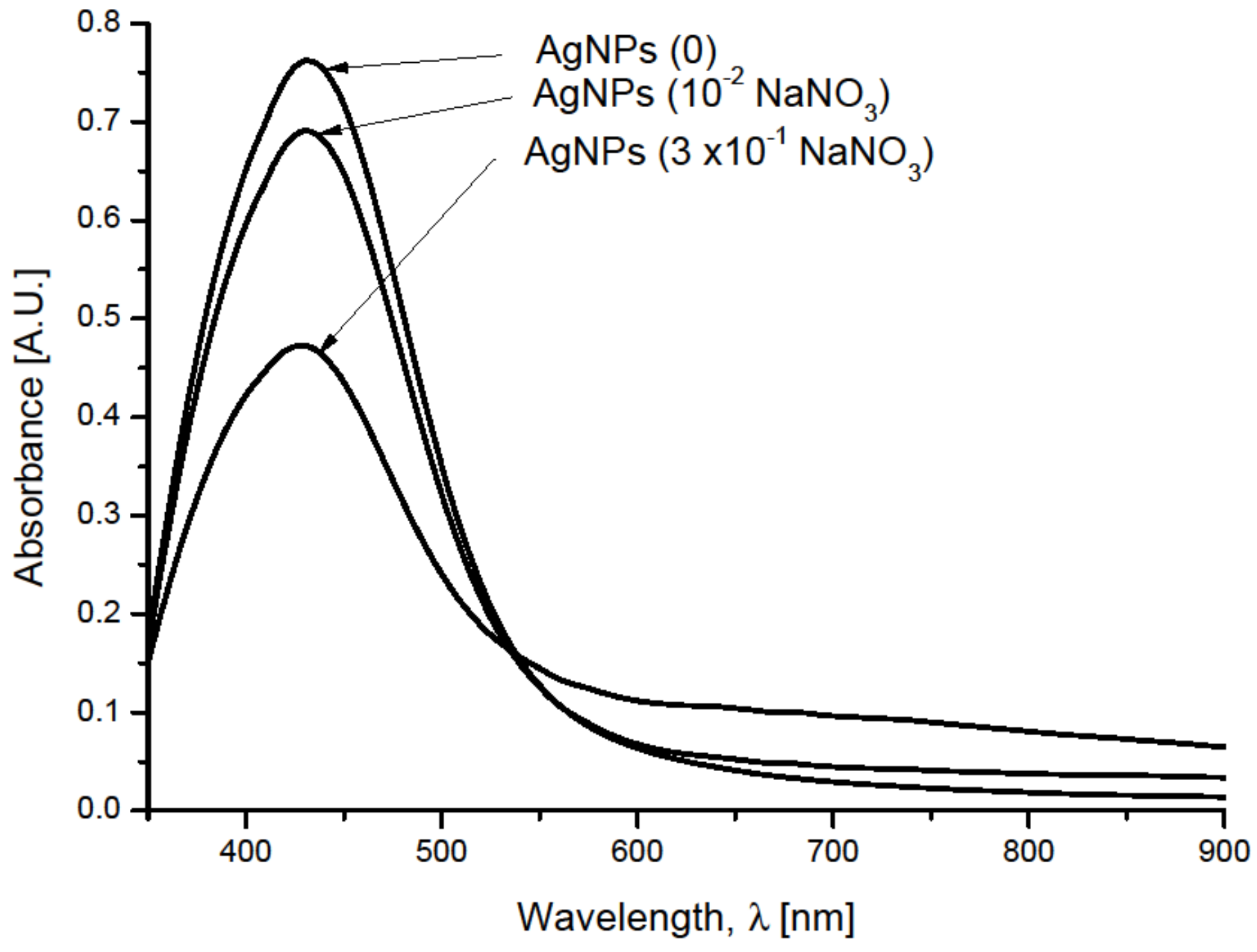

Figure 05

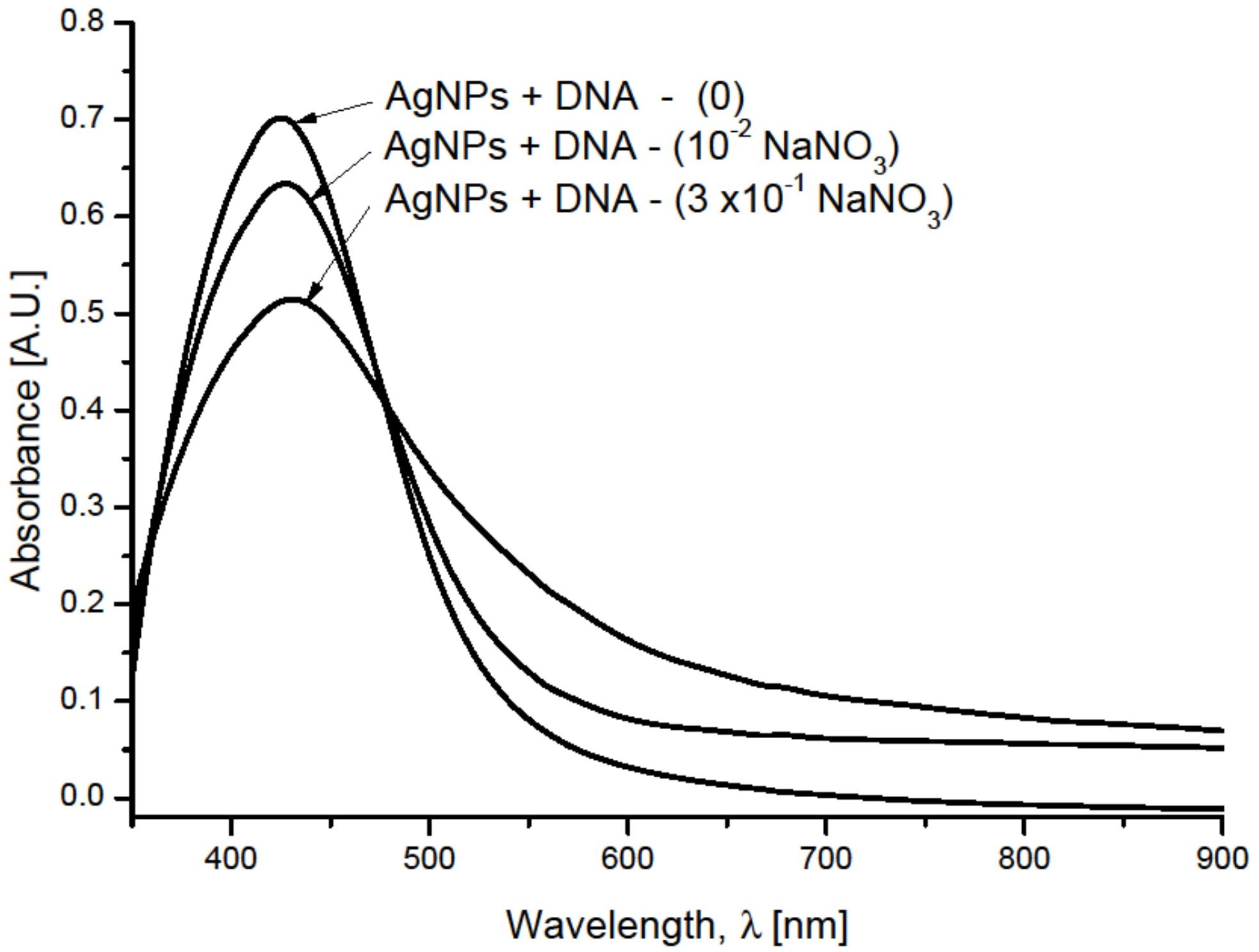

Figure06

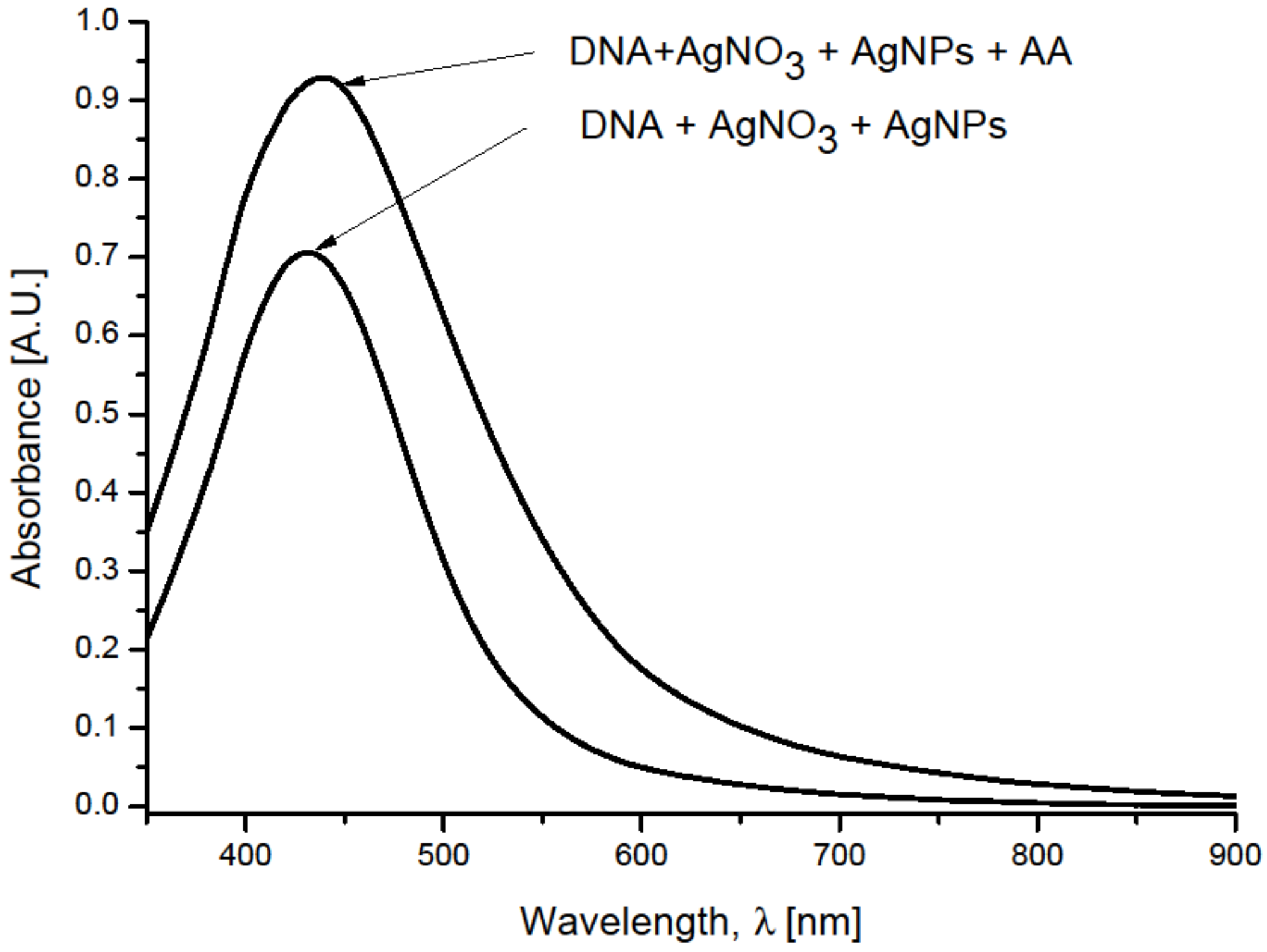

Figure07

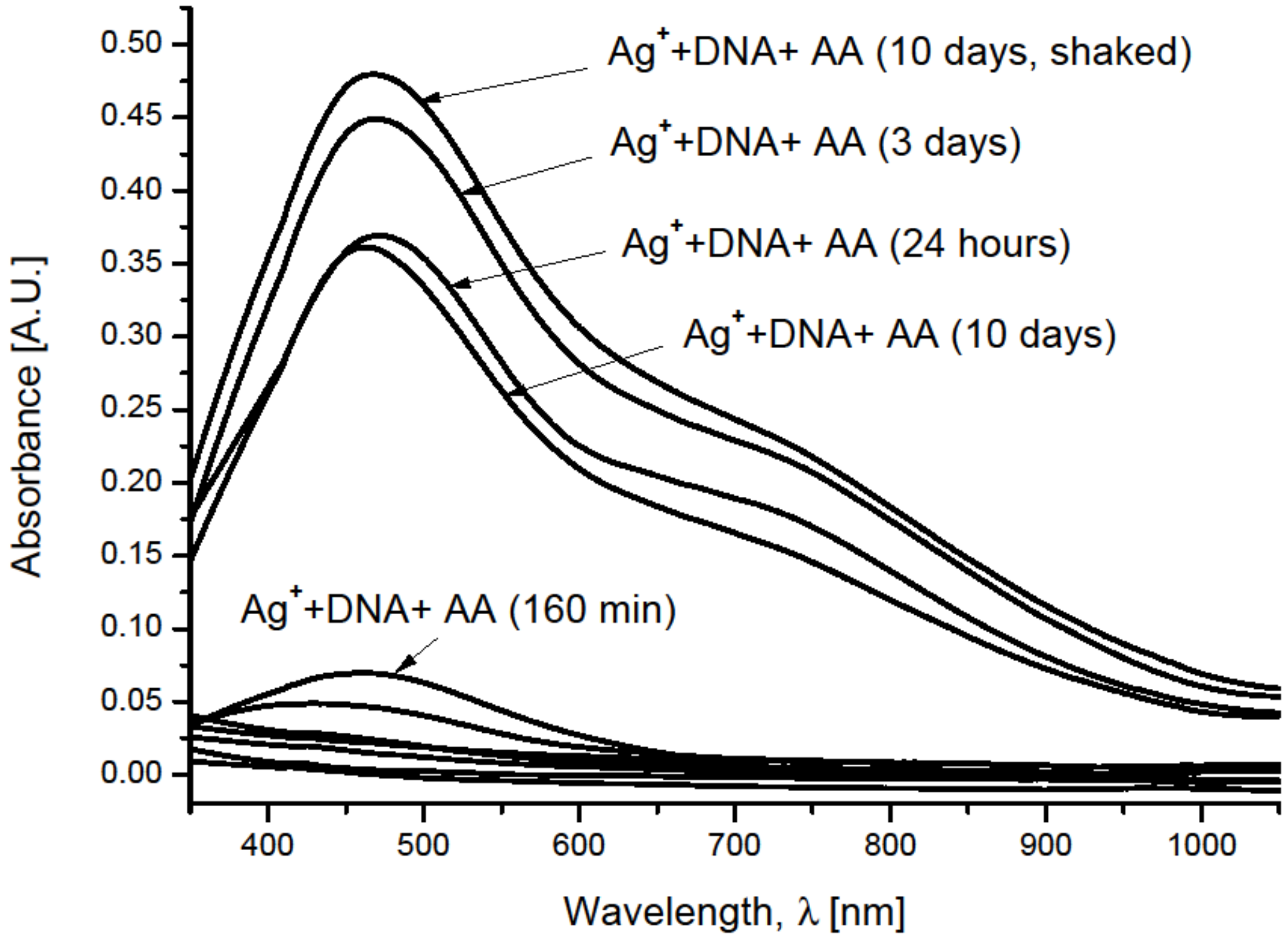

Figure 08

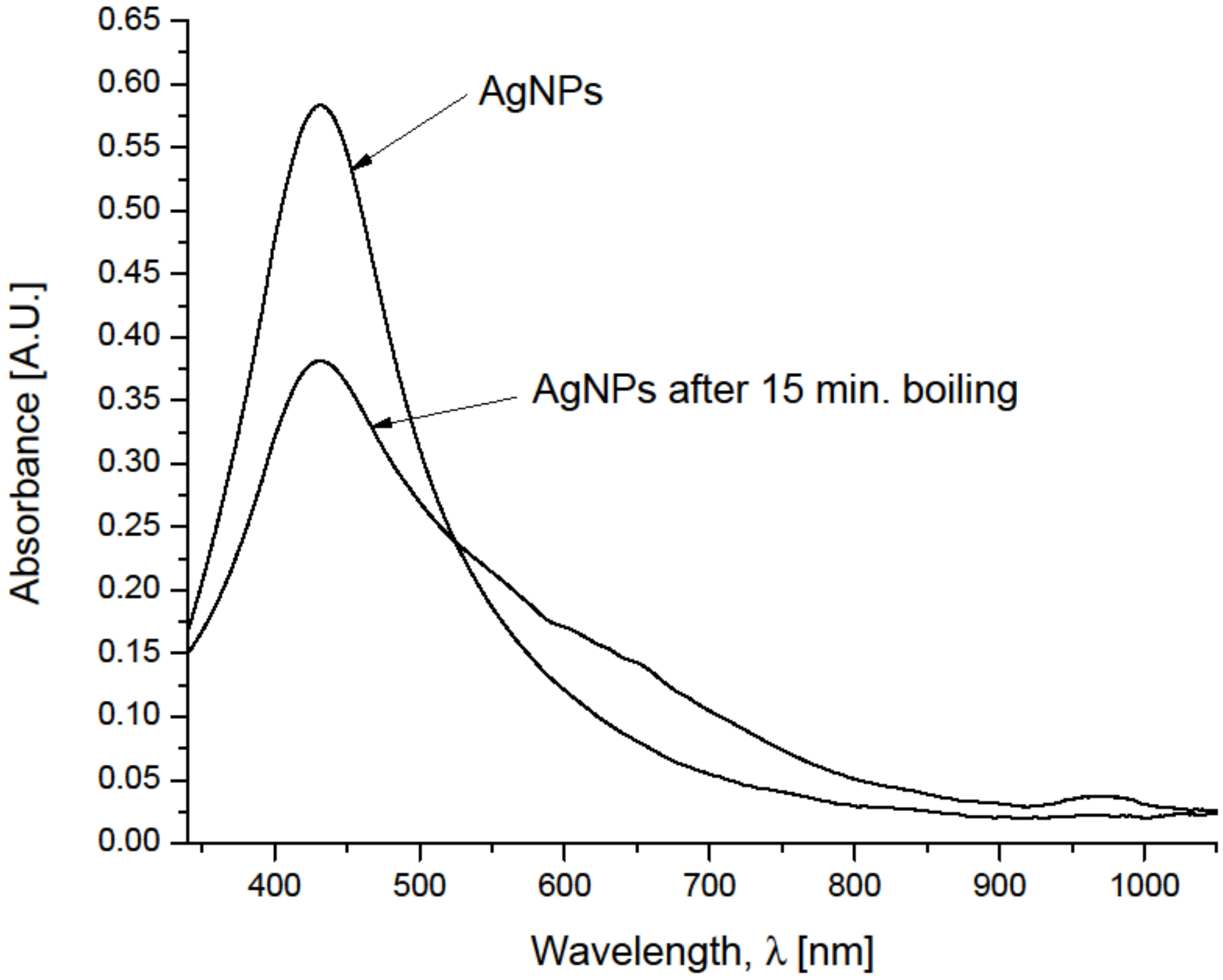

Figure09

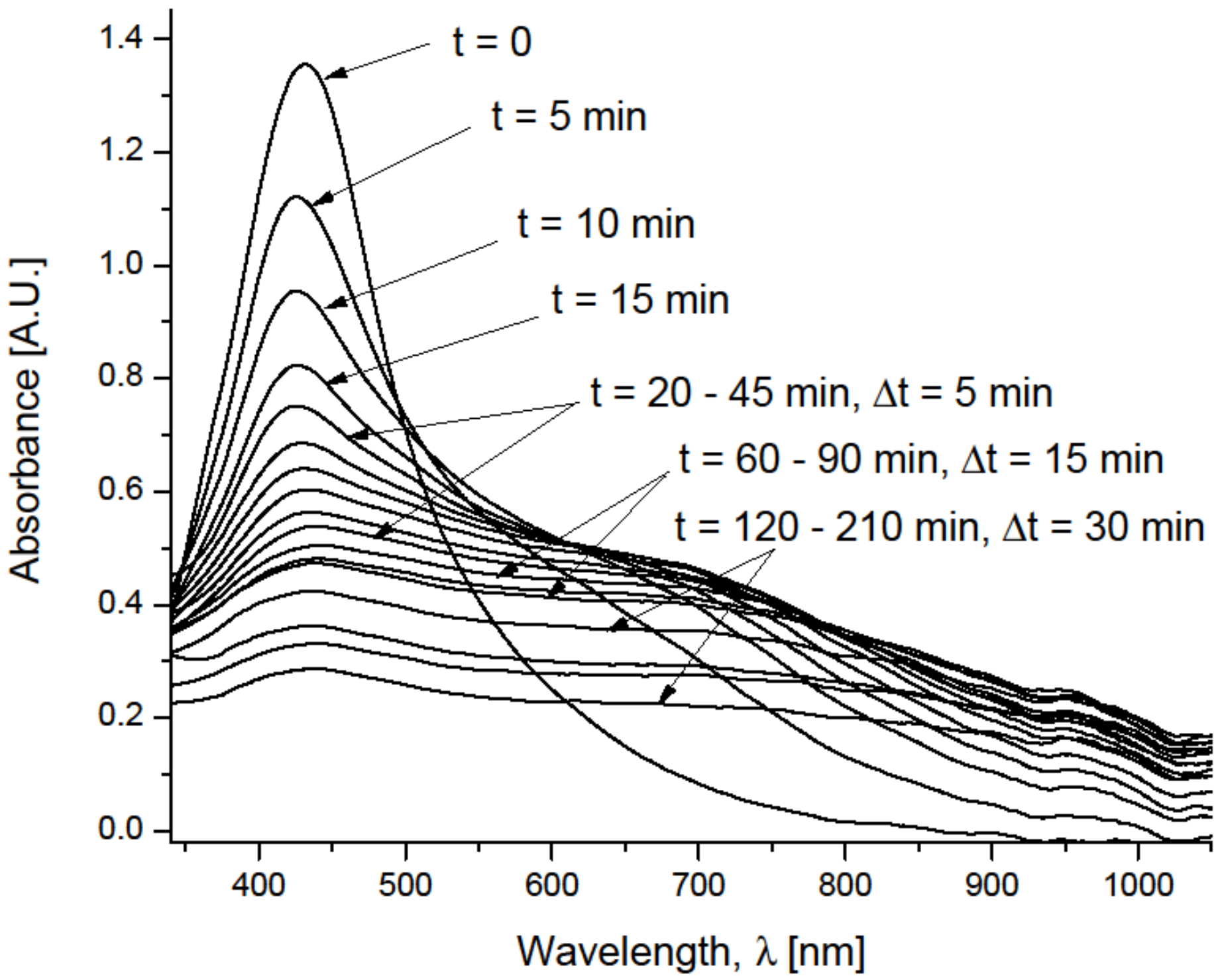

Figure10

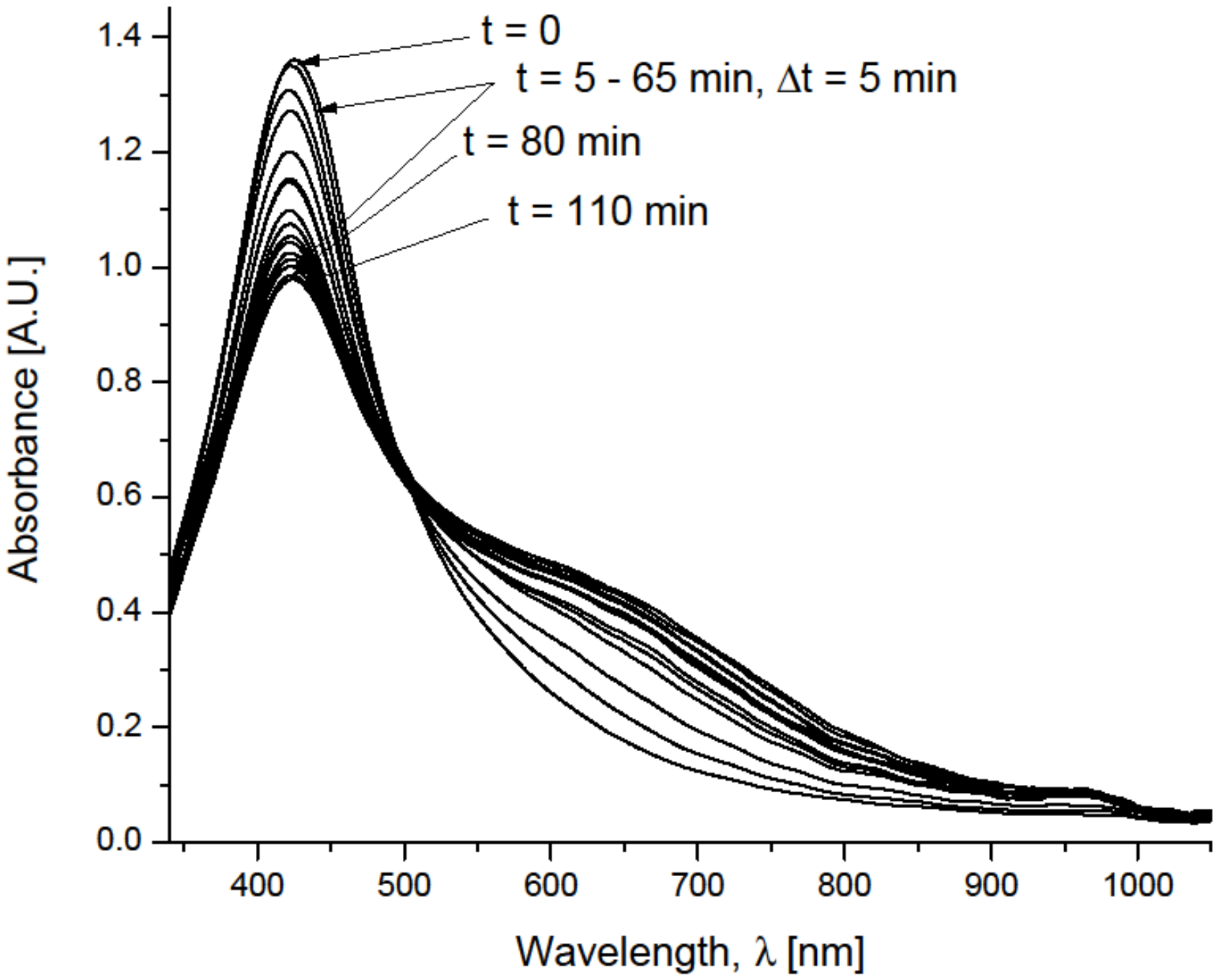

Figure11

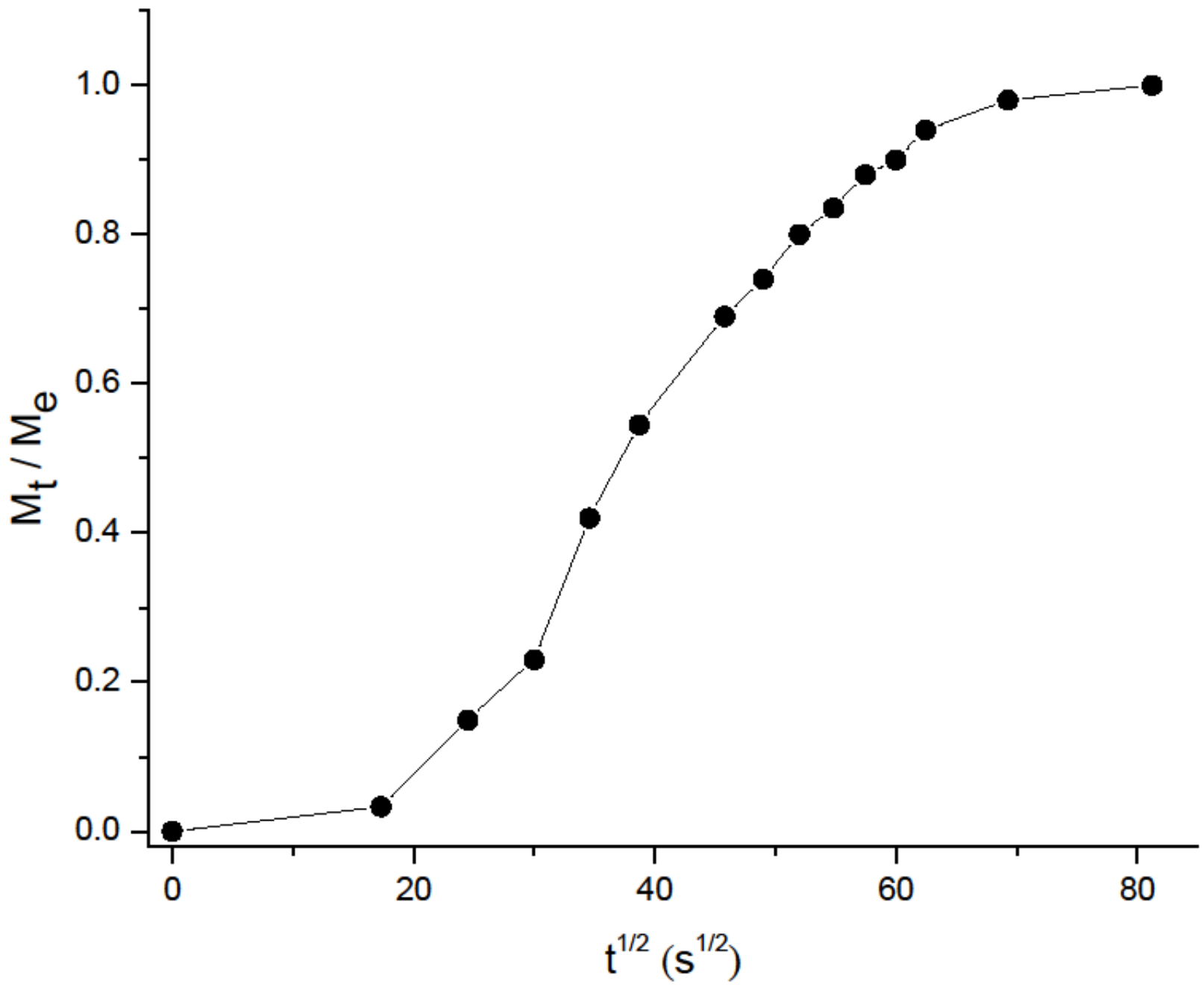

Figure12

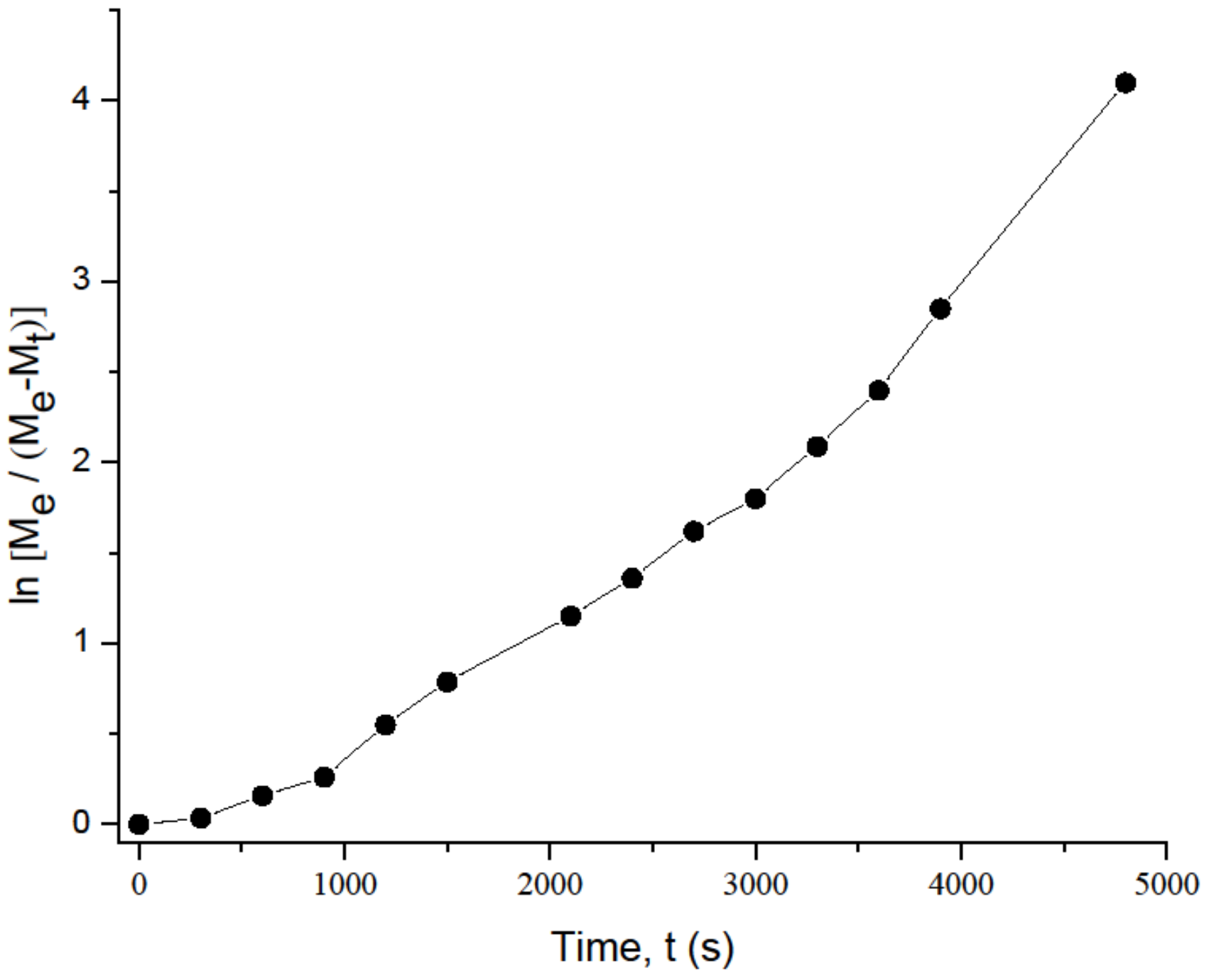

Figure 13

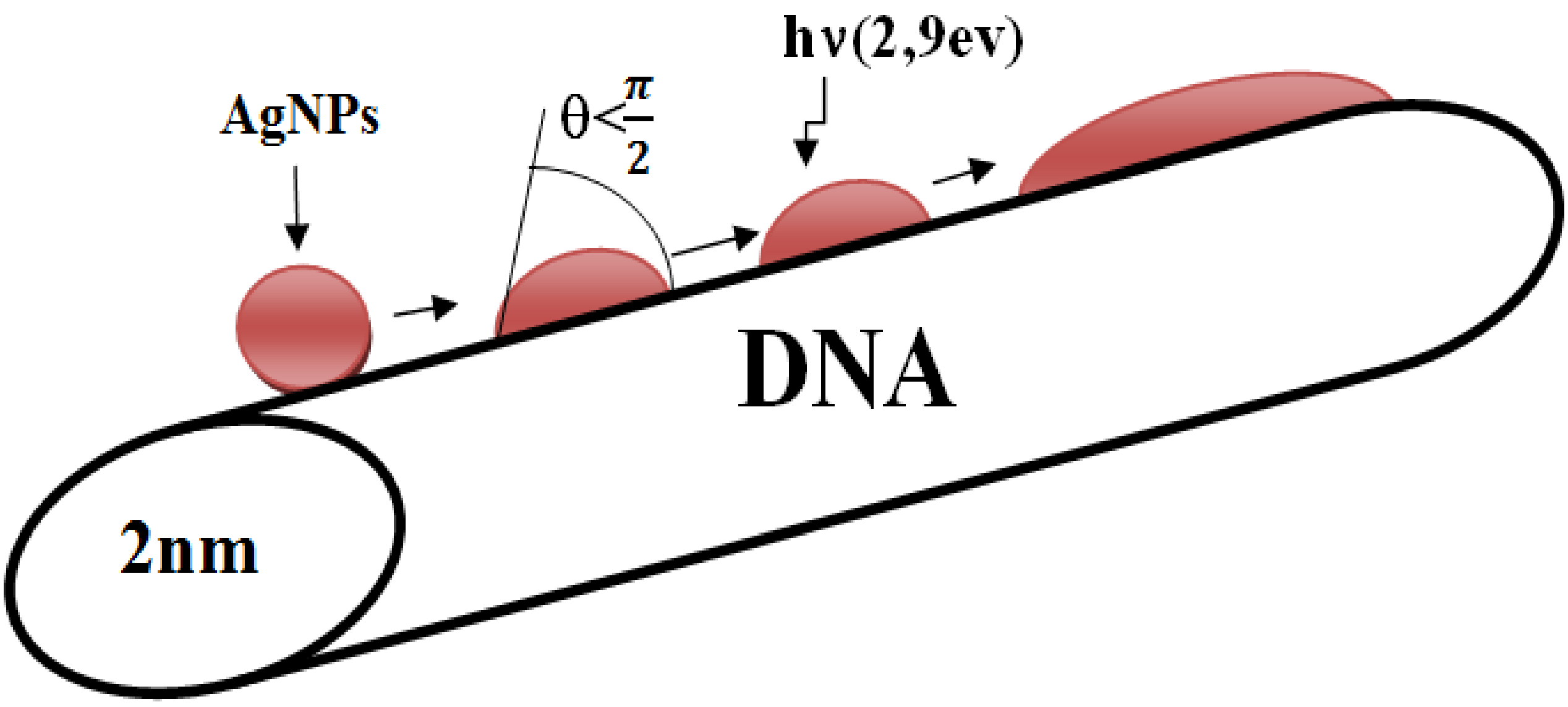

Figure14

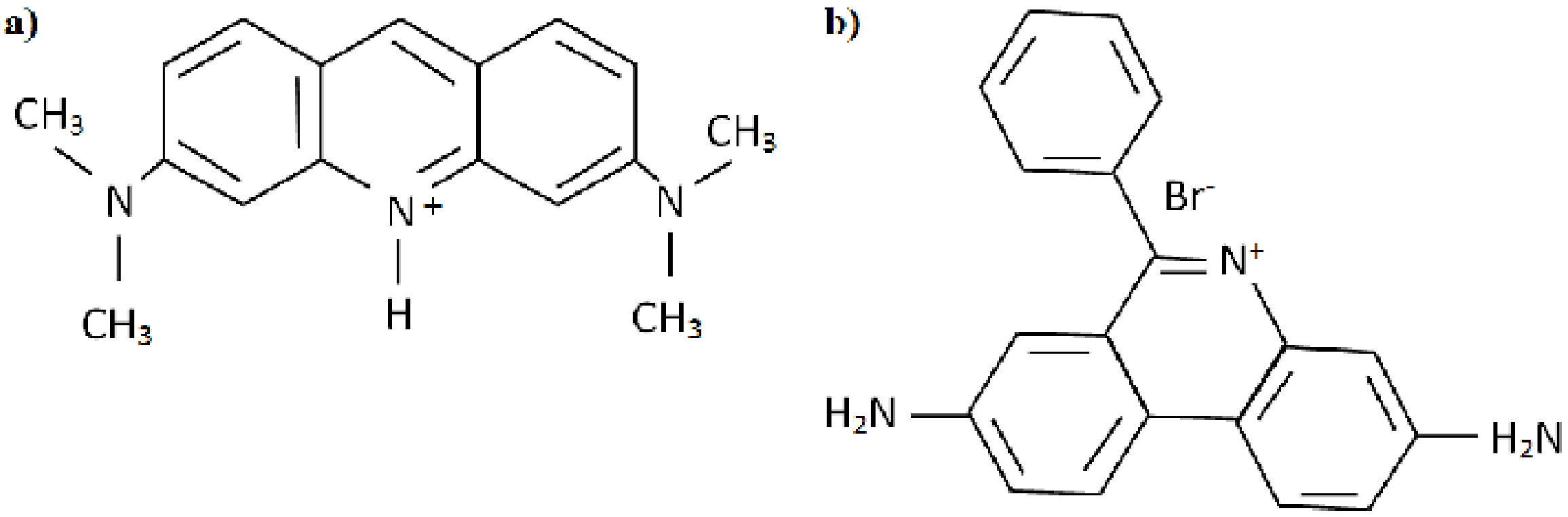

Figure15

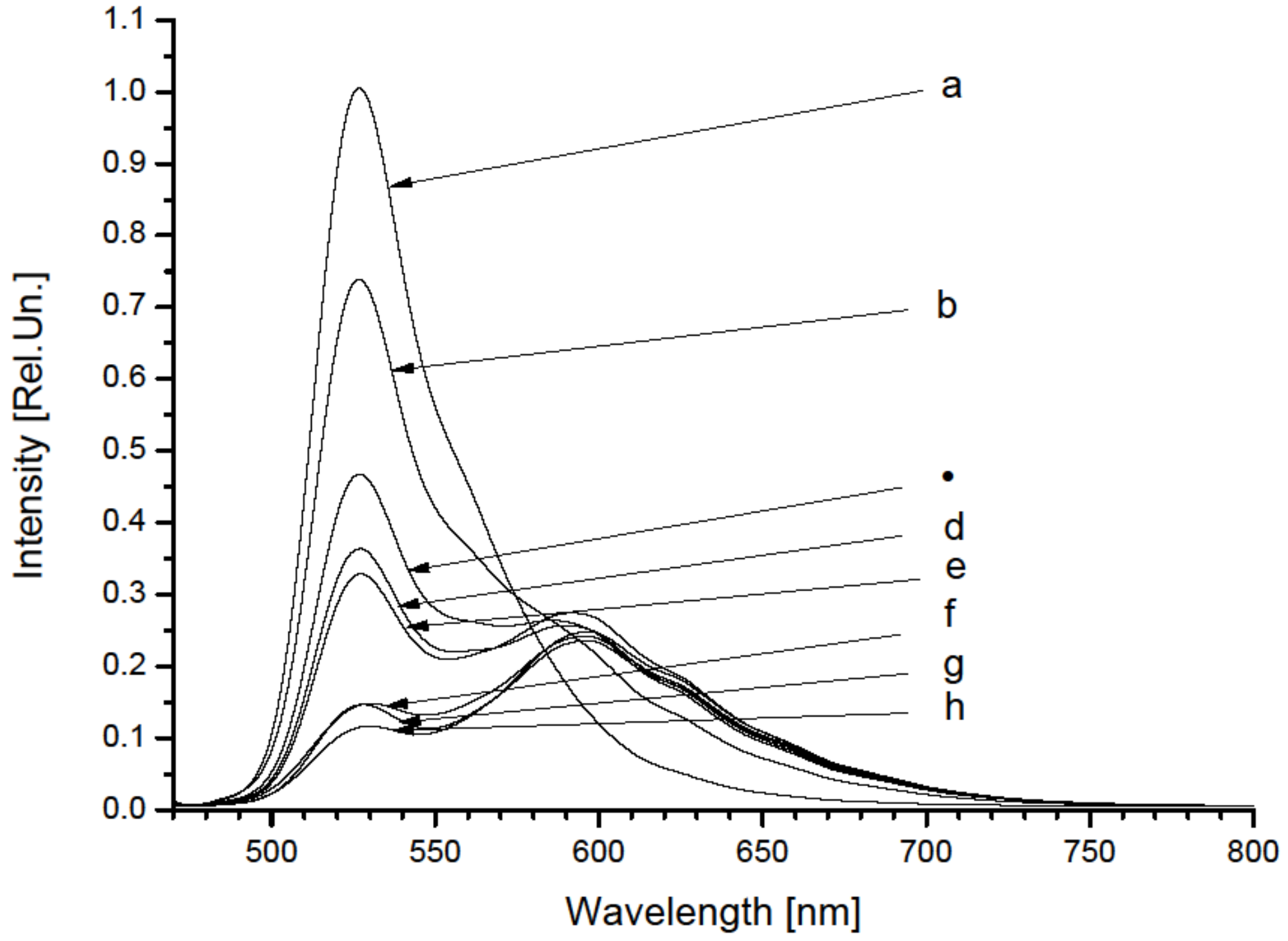

Figure16

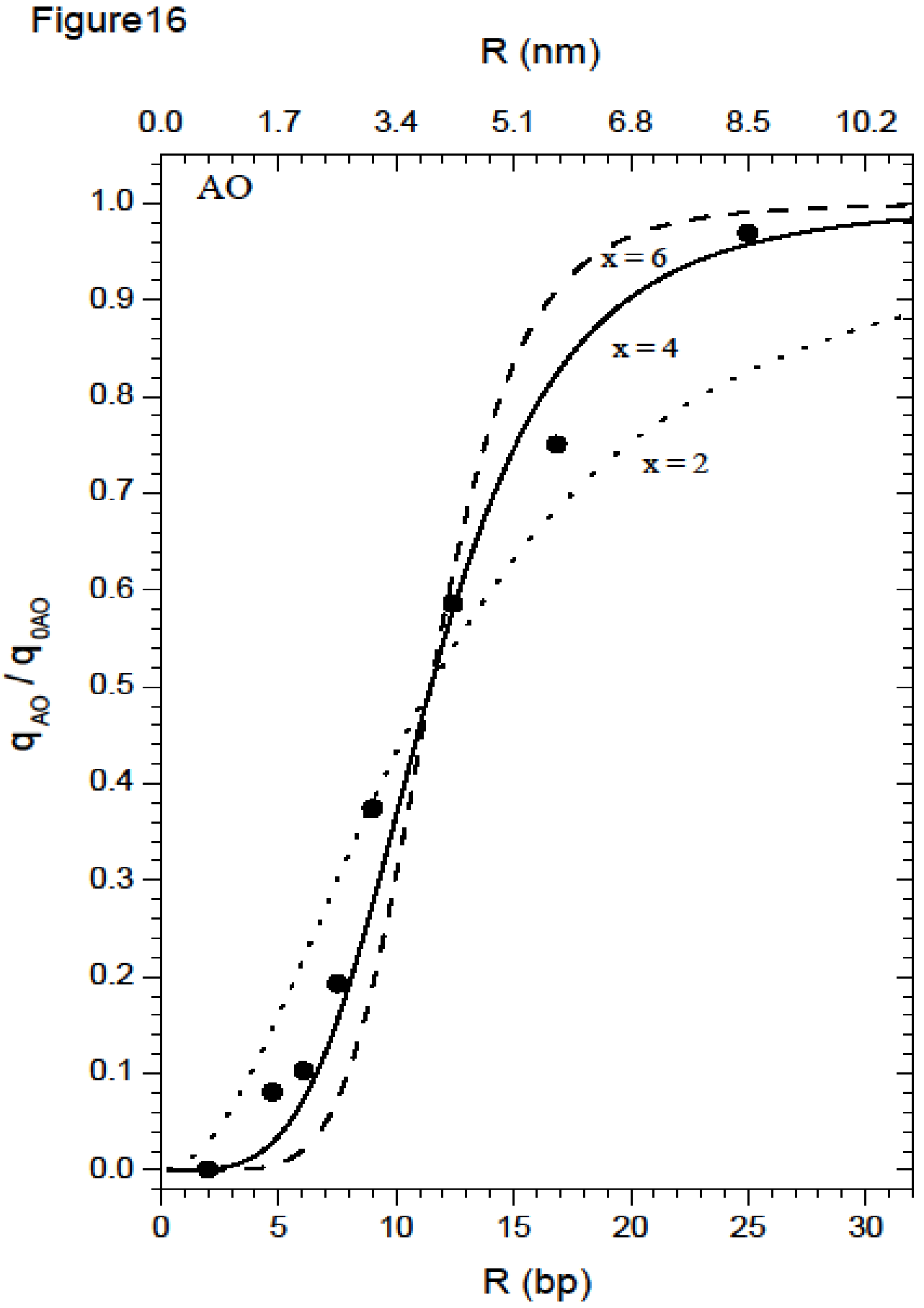

Figure17

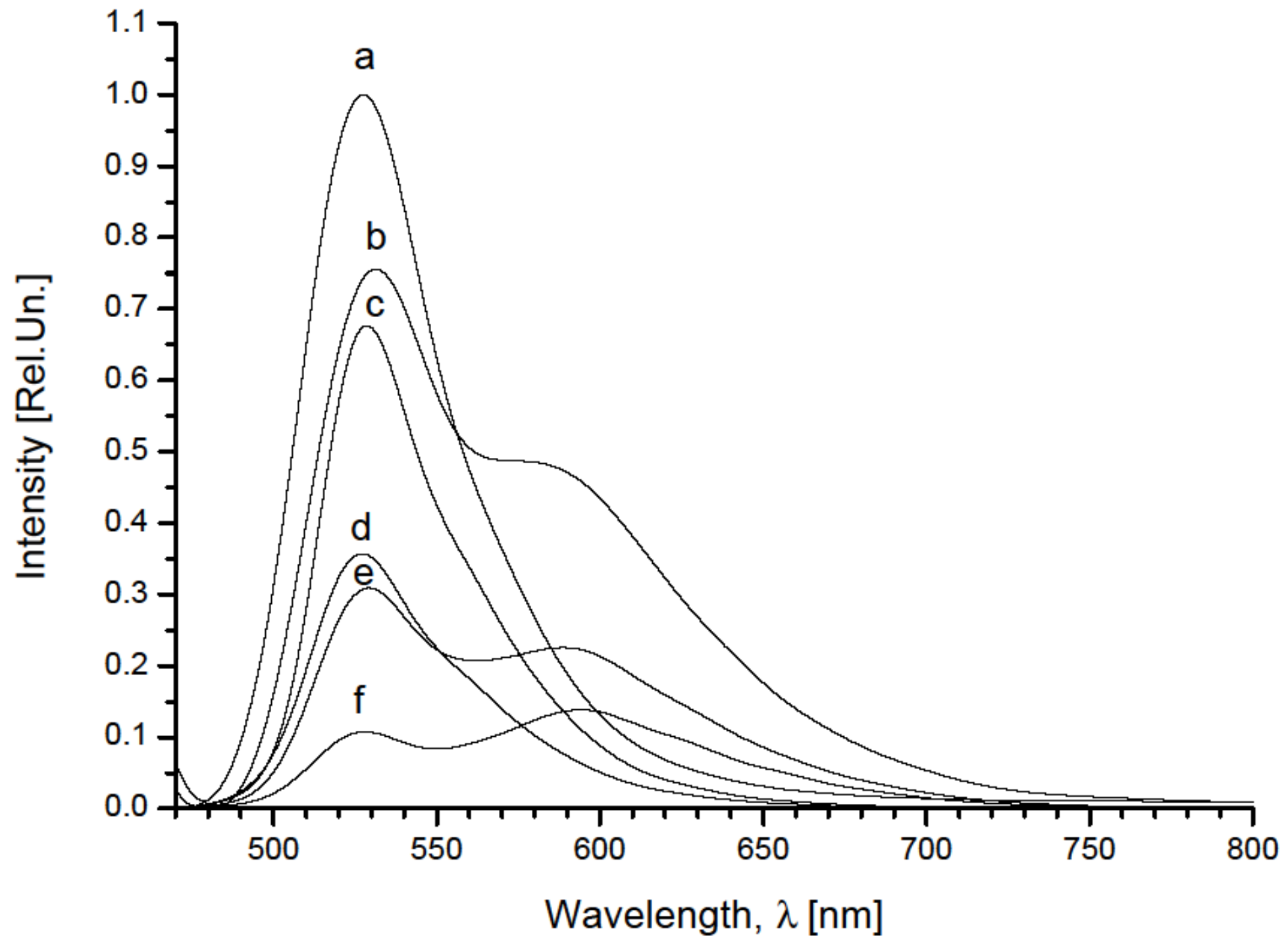

Figure18

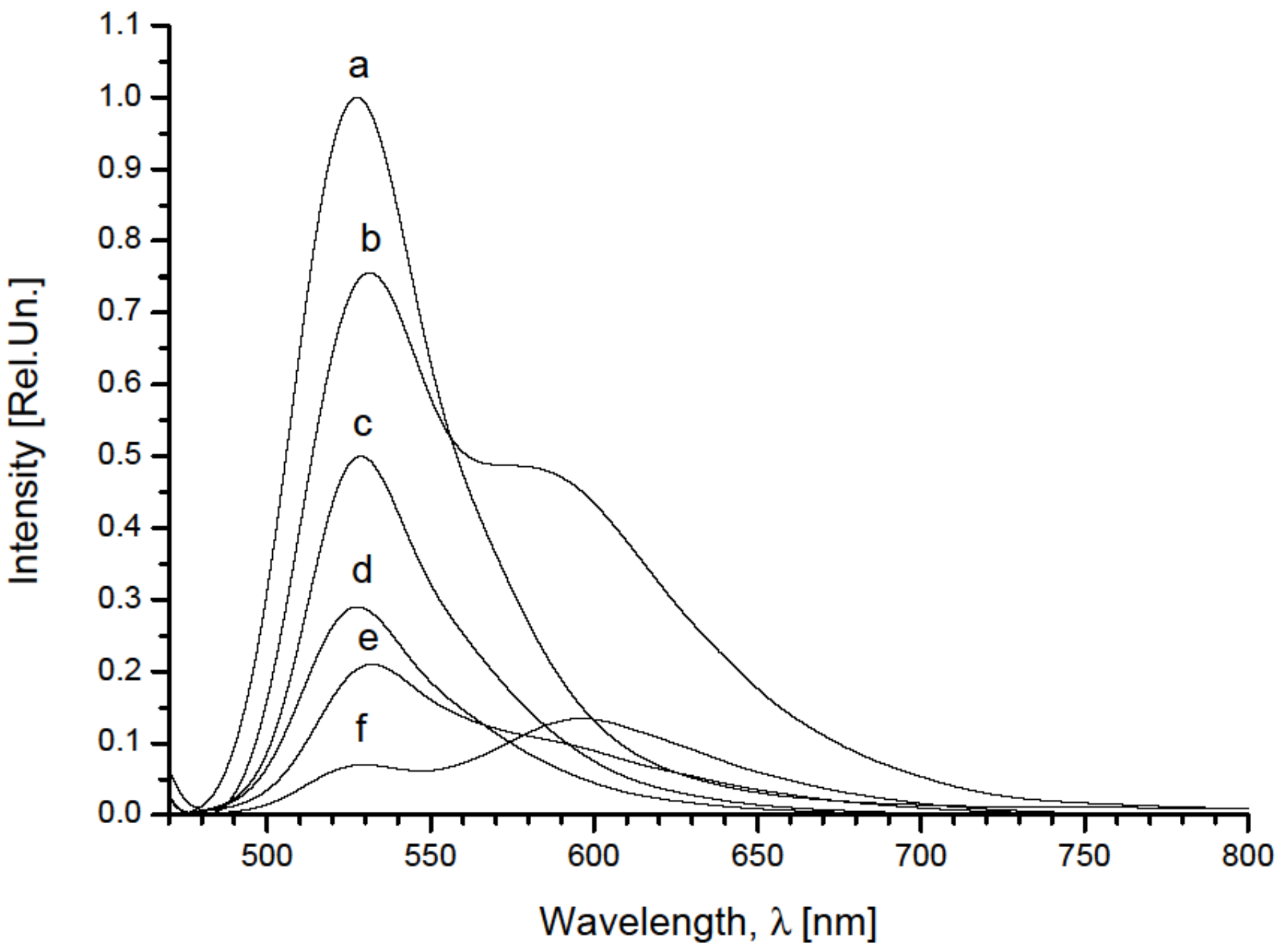

Figure19

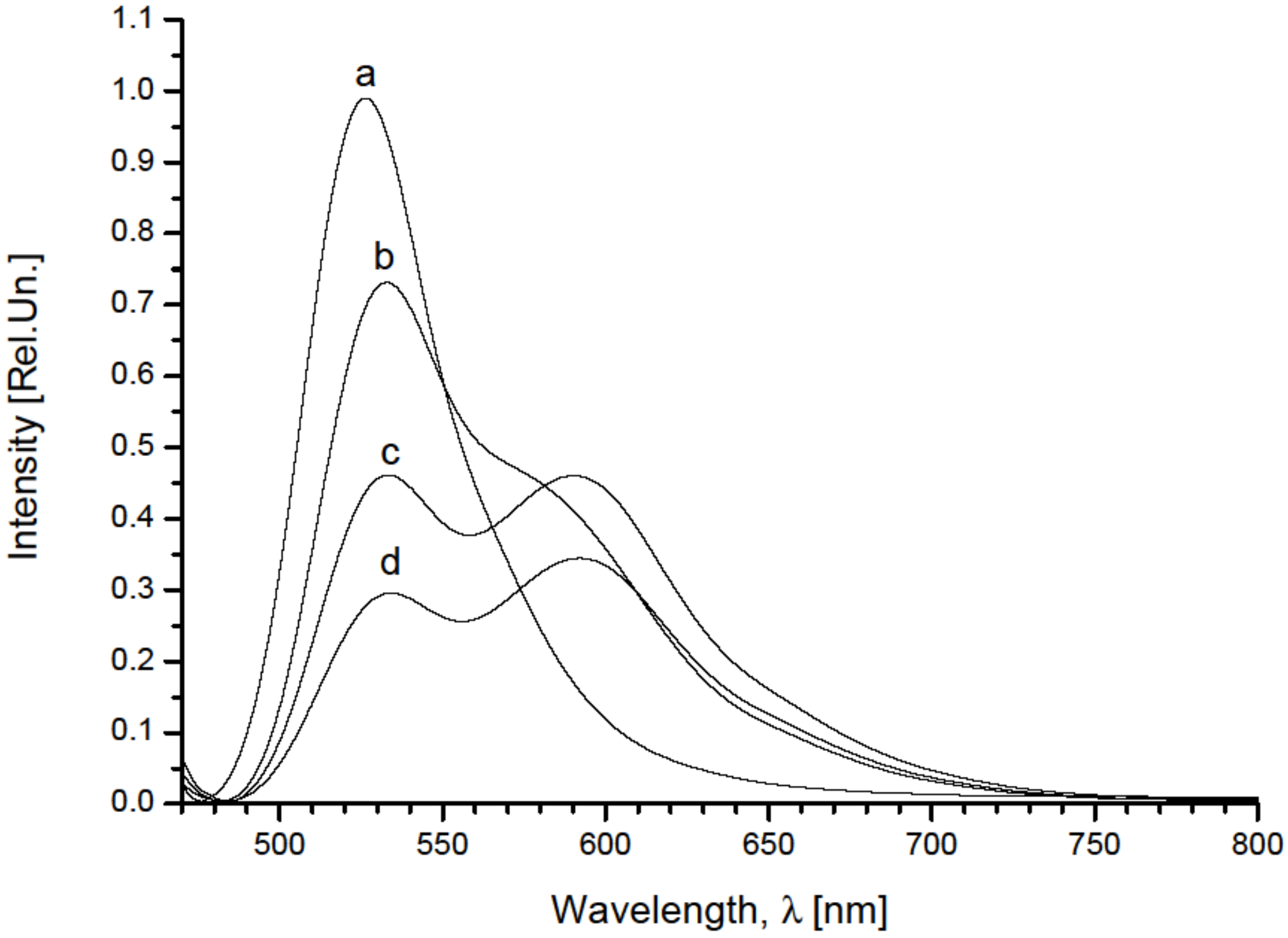

Figure20

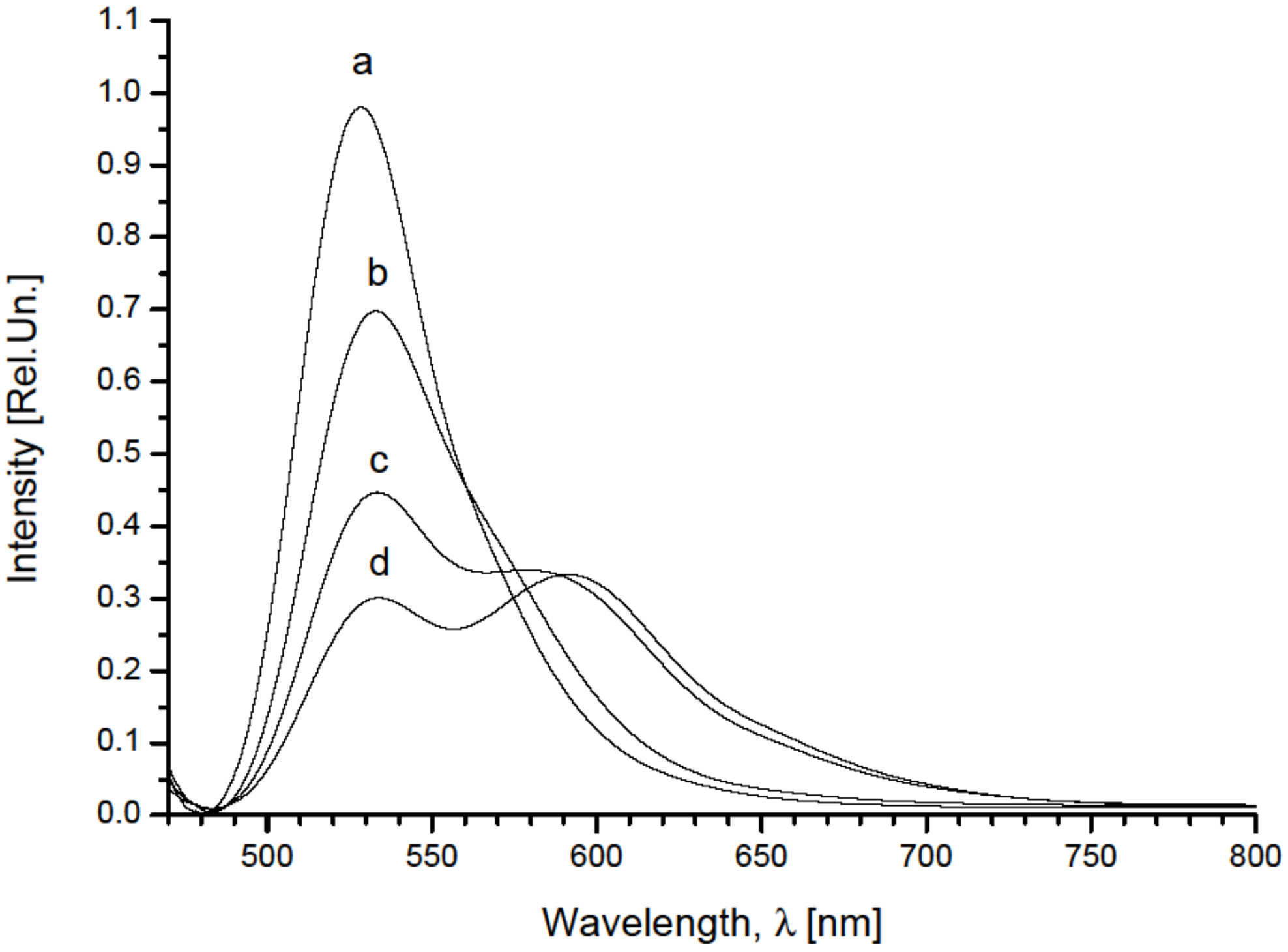

Figure21

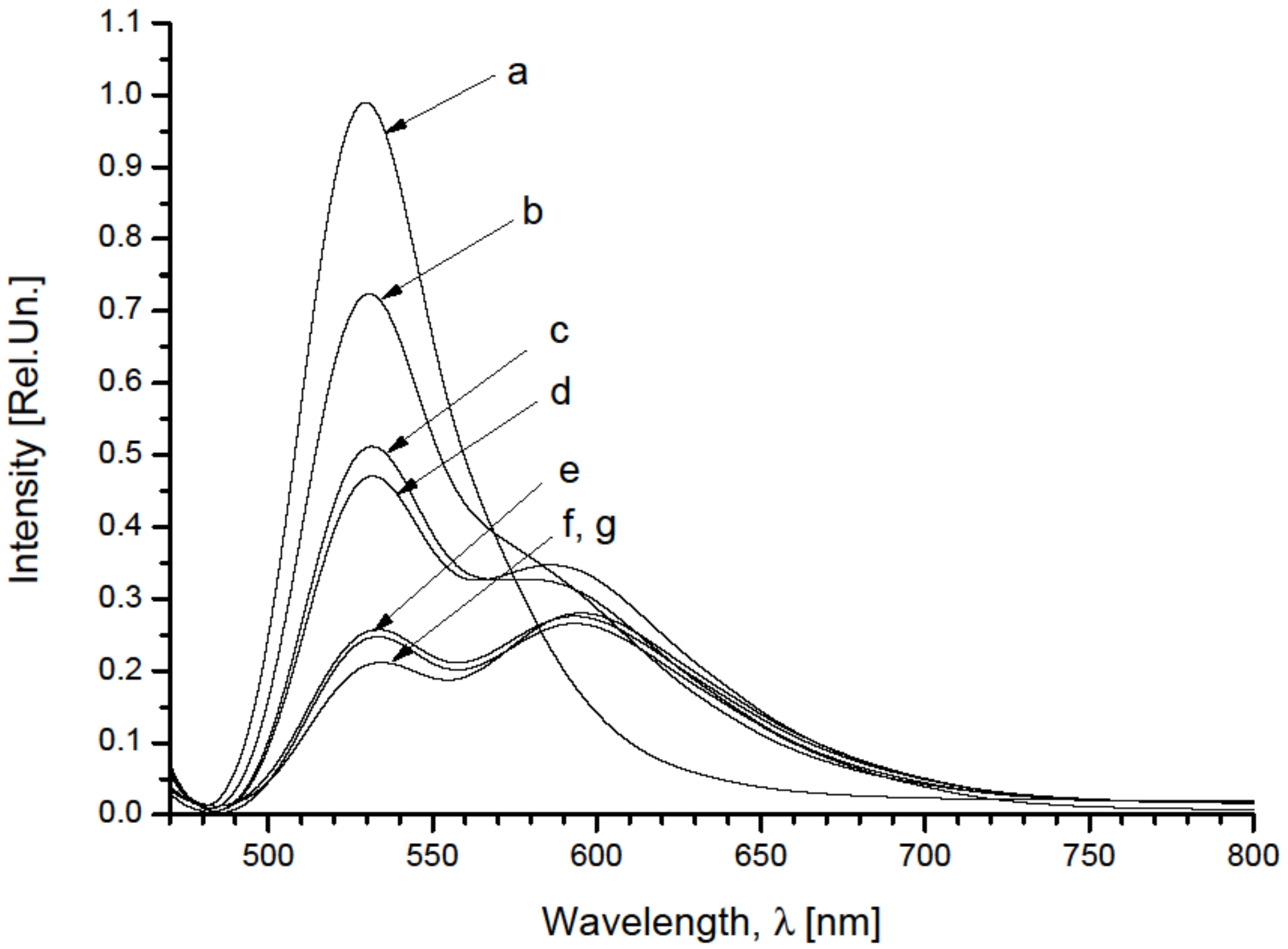

Figure22

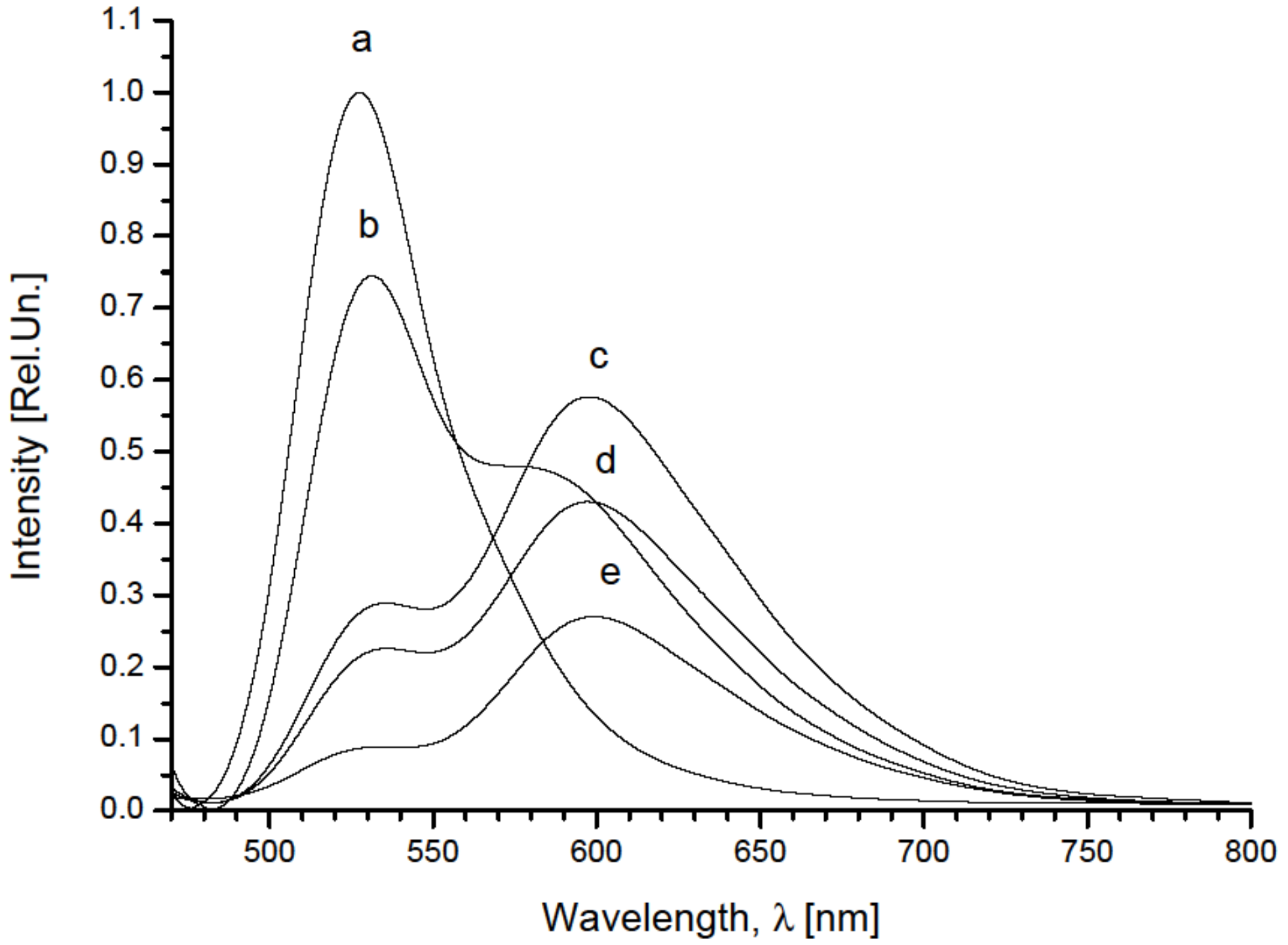

Figure23

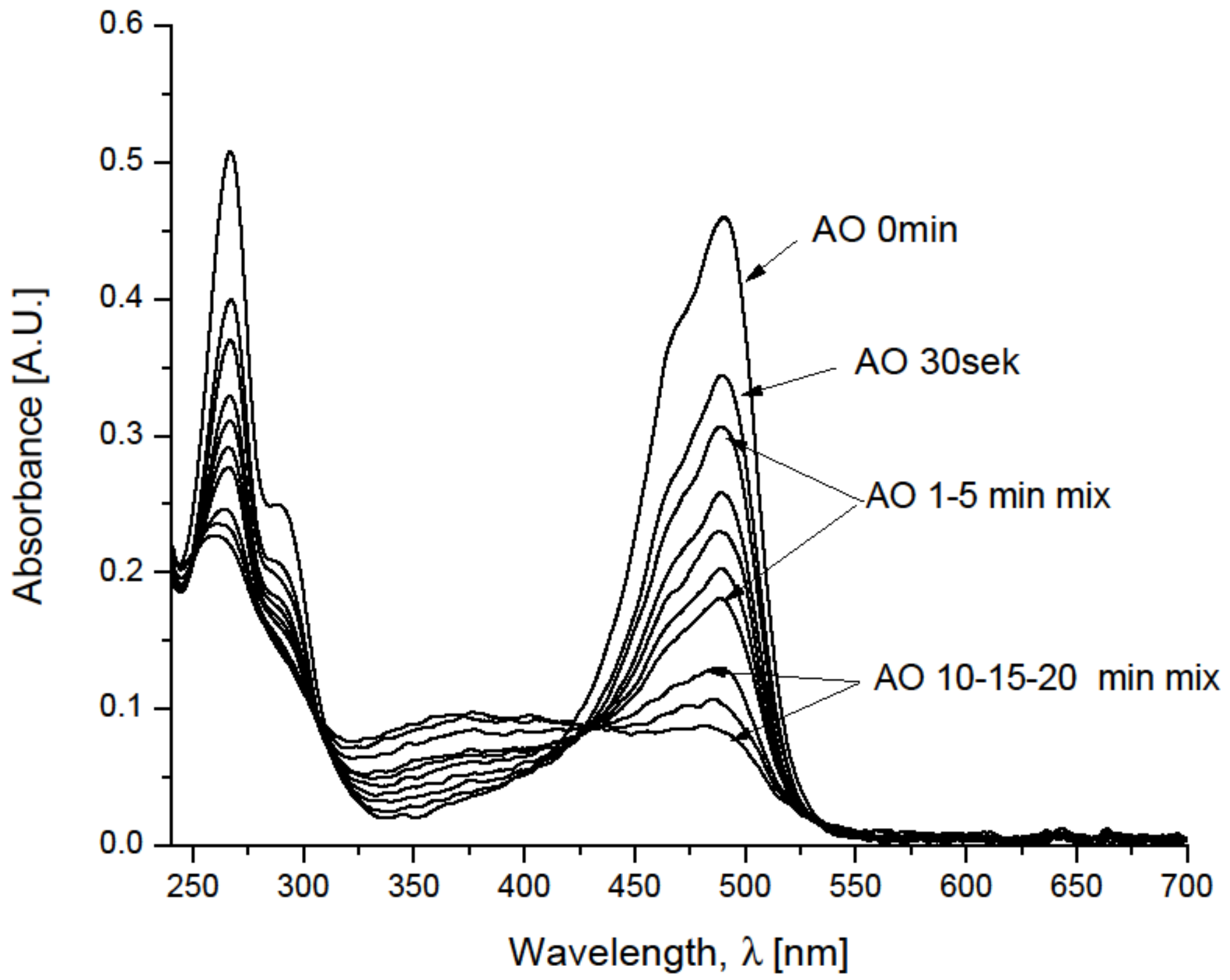

Figure24

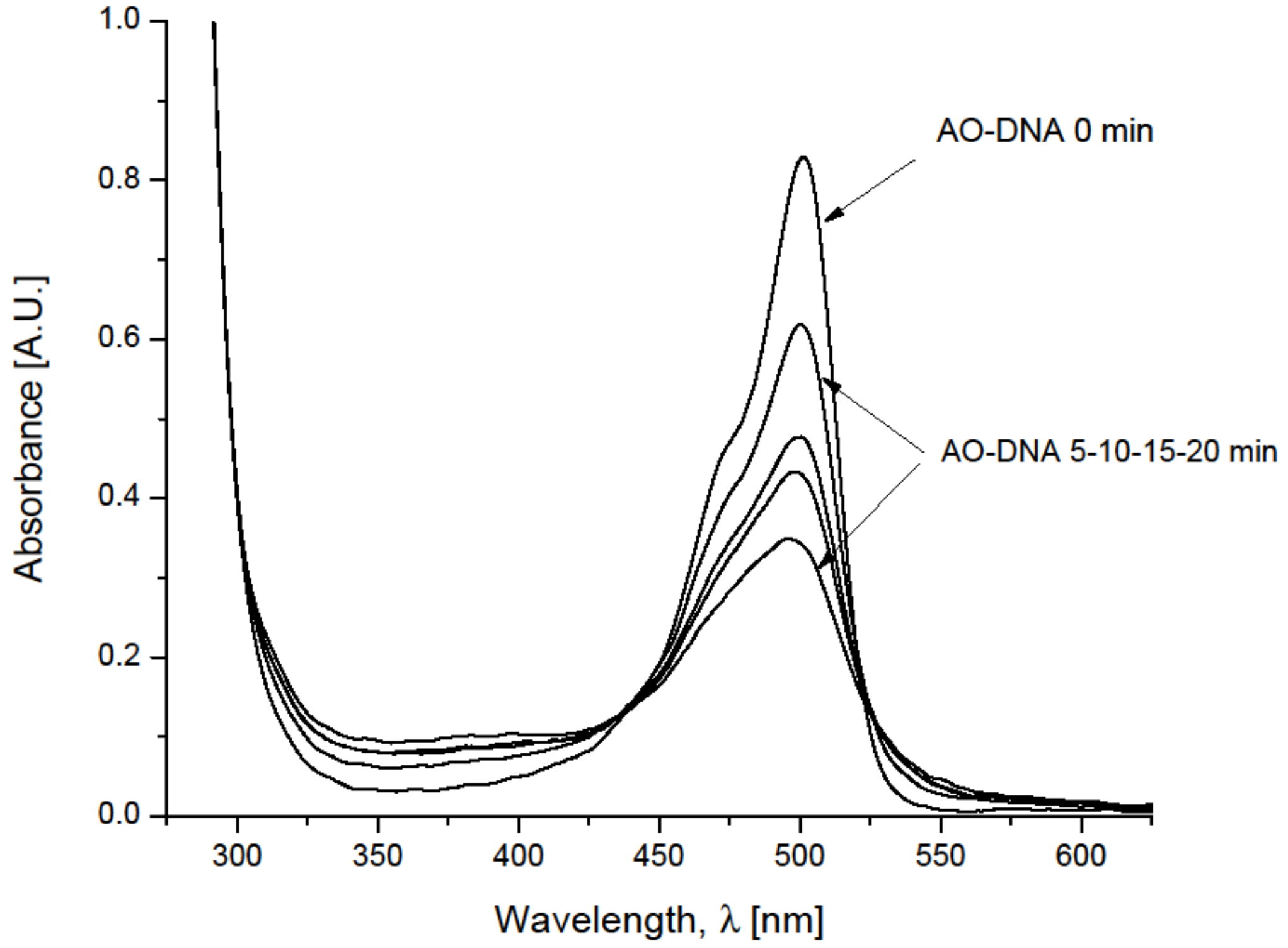

Figure25

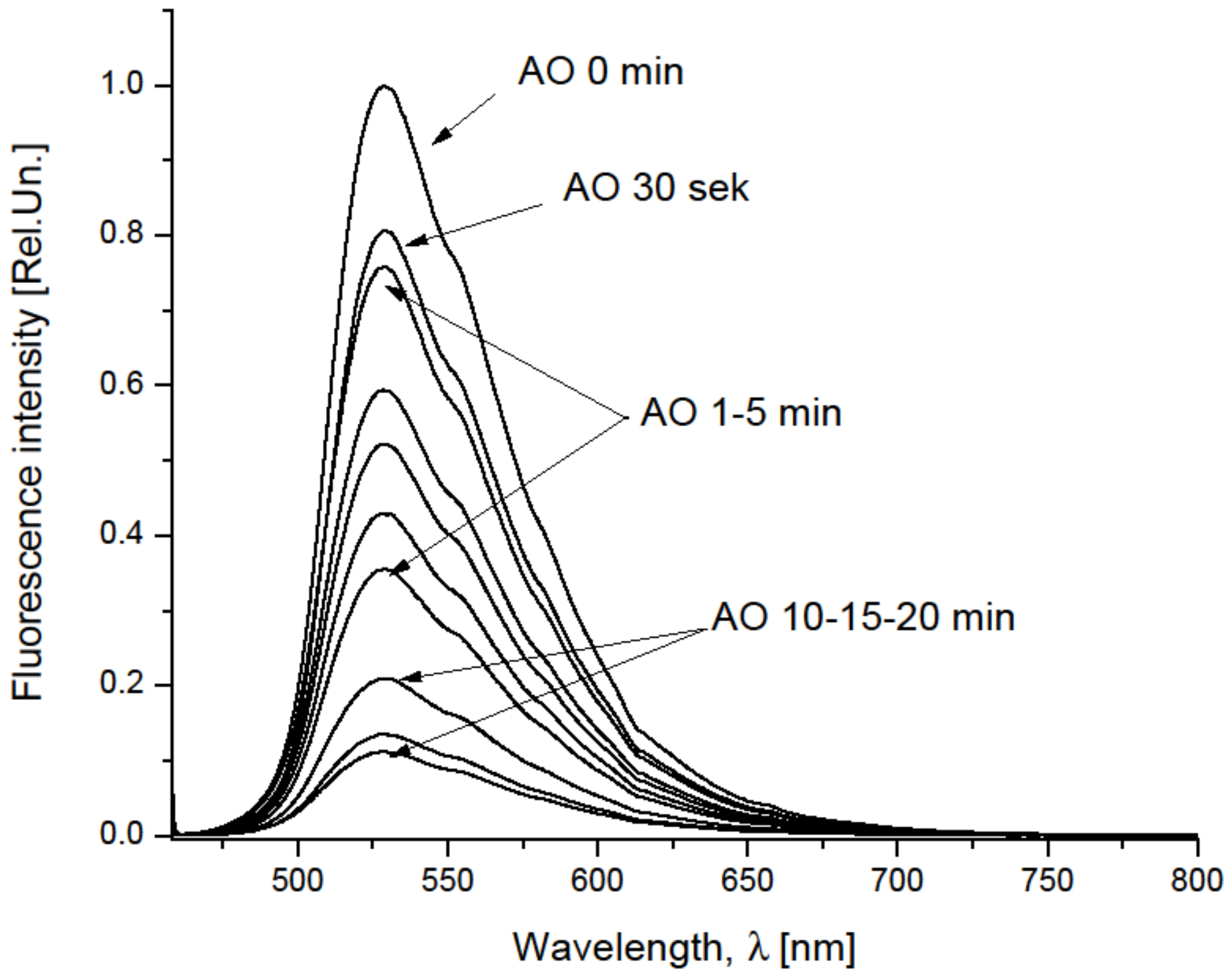

Figure26

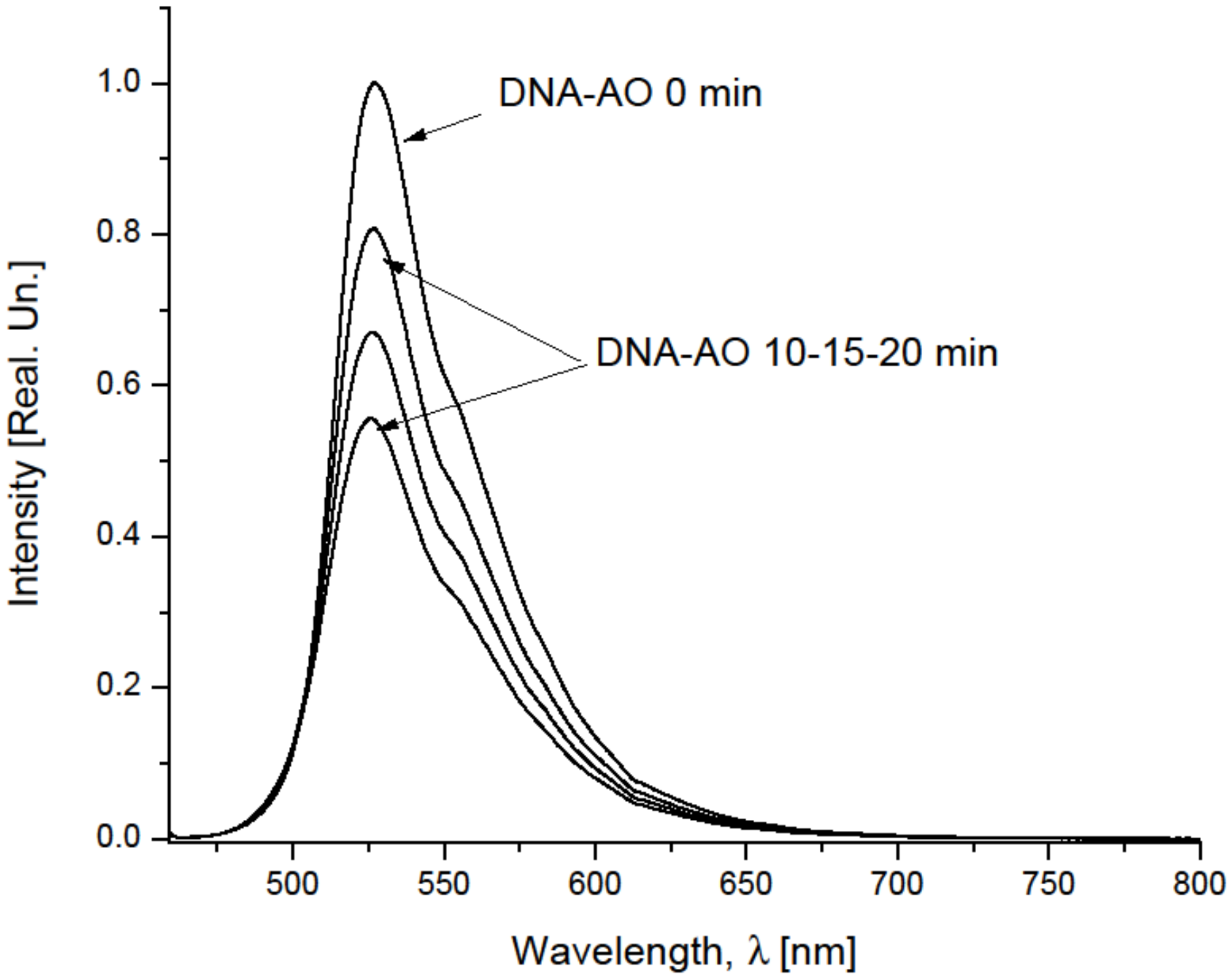

Figure27

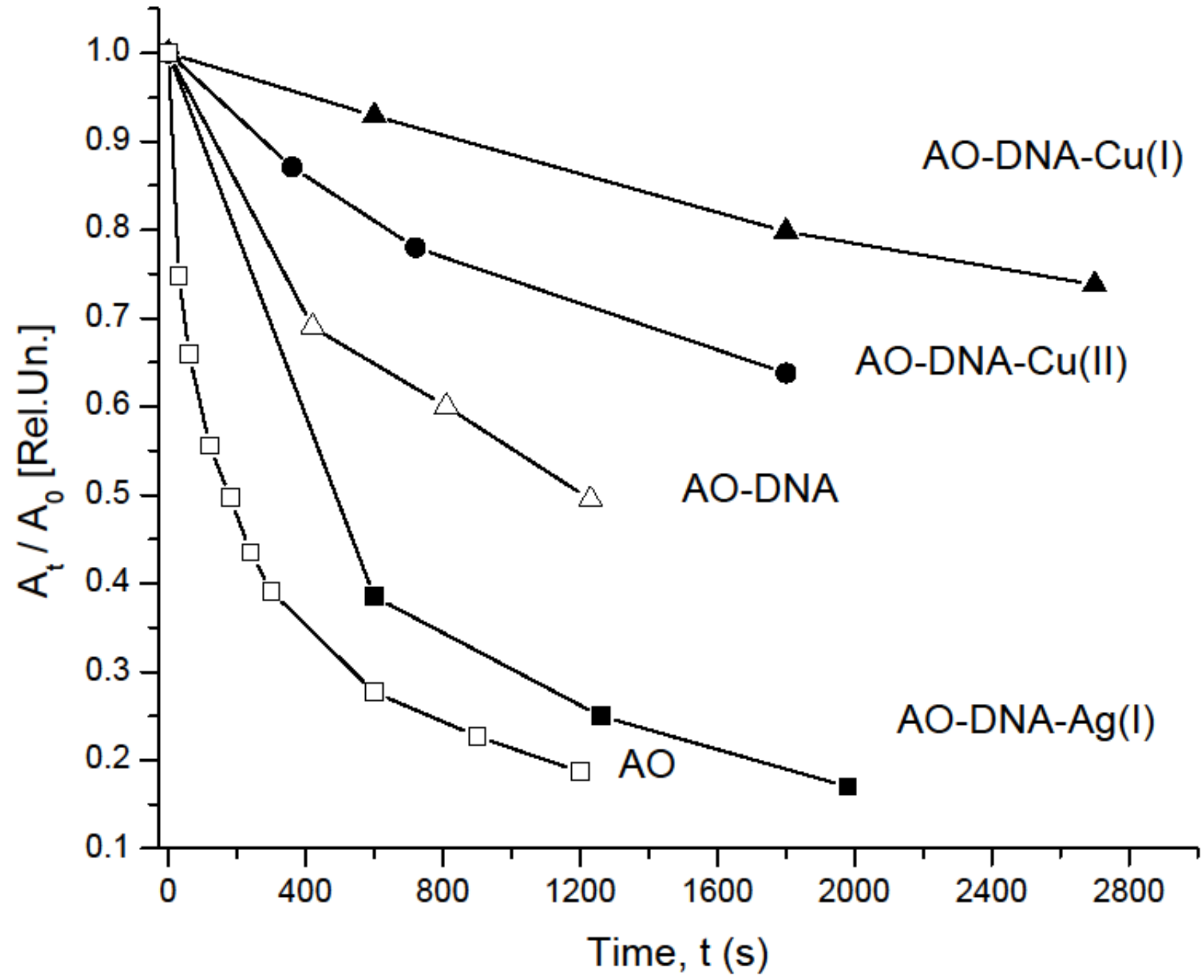

Figure28

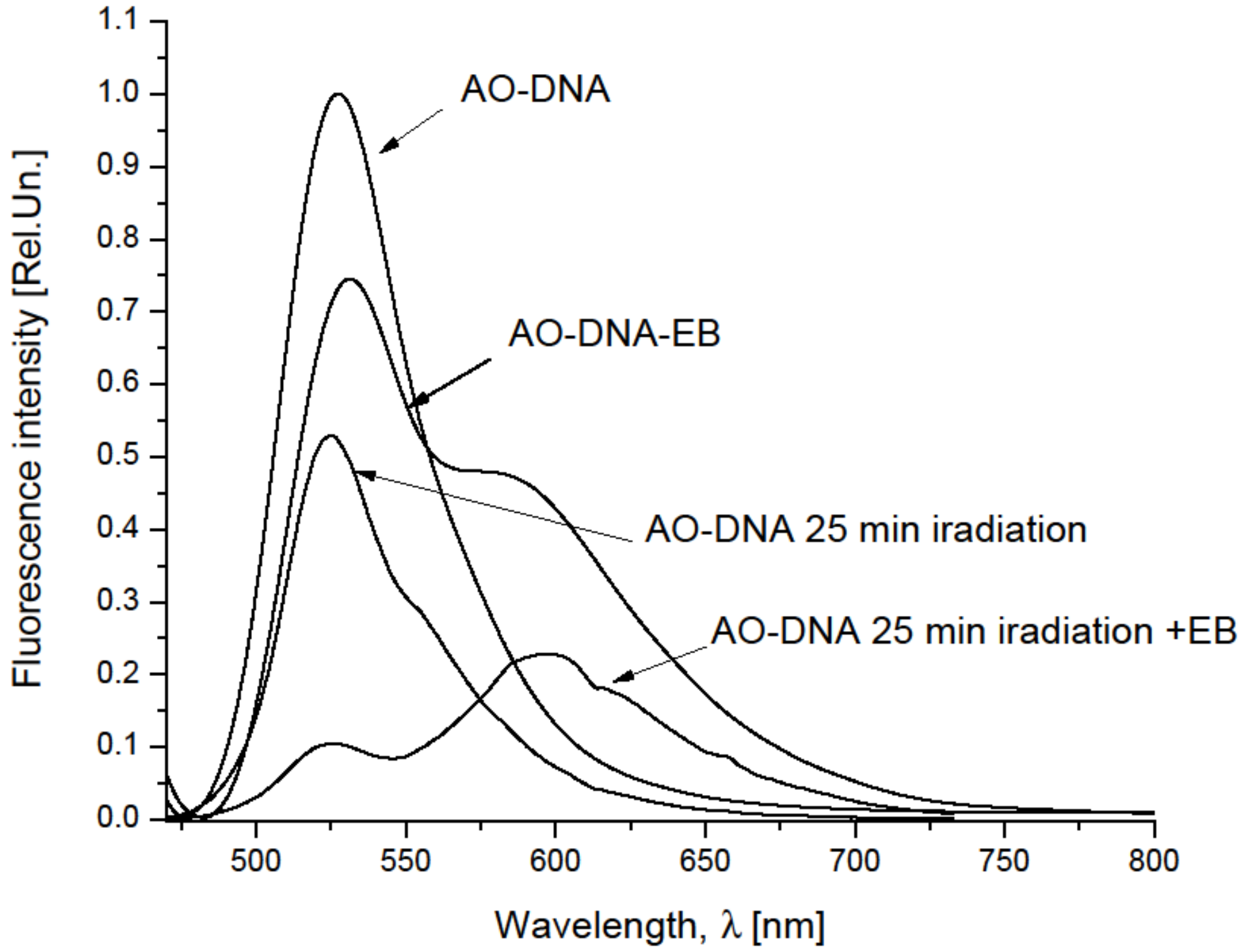

**Figure 29**

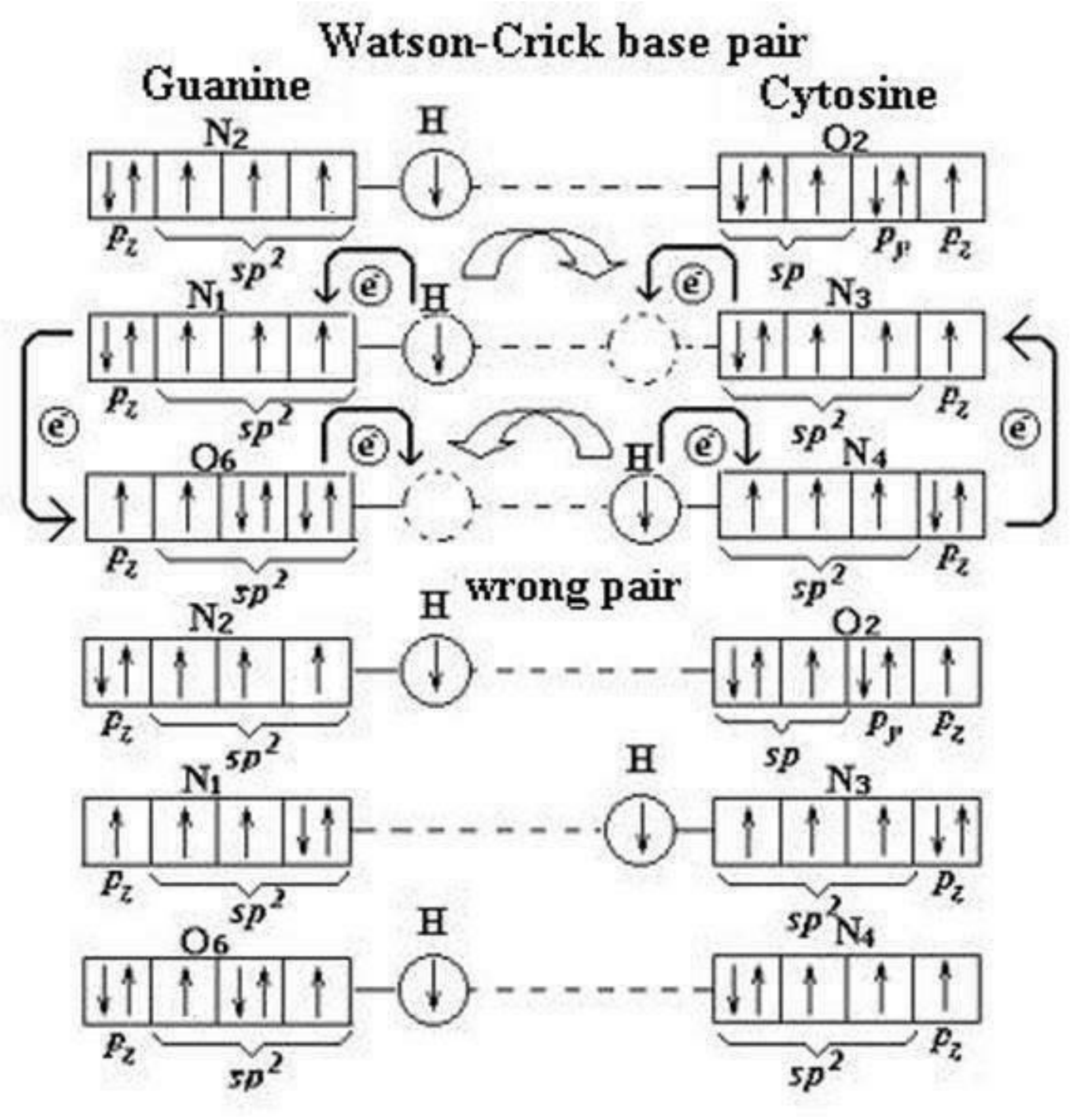

**Figure 30**

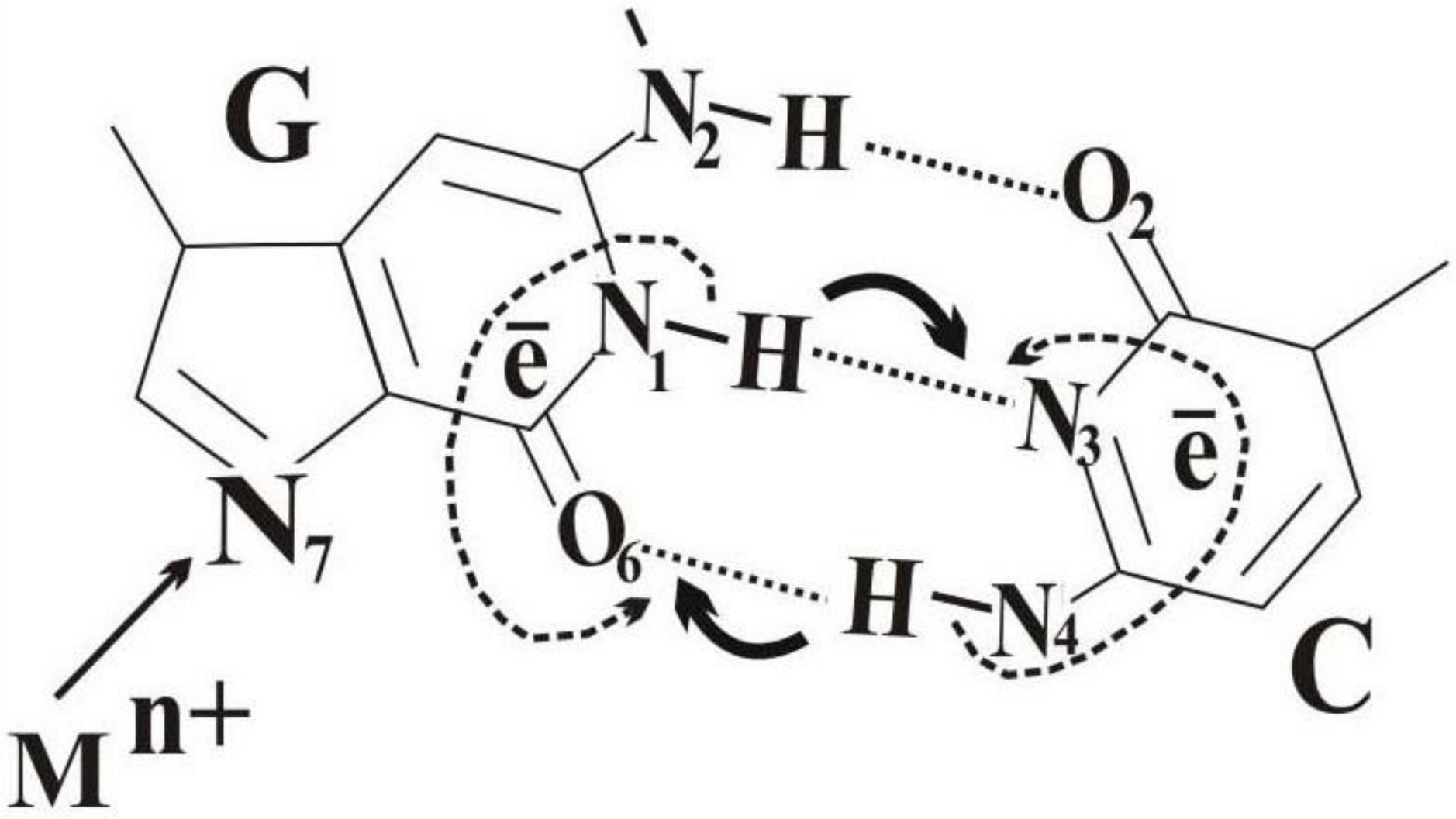

**Figure 31**

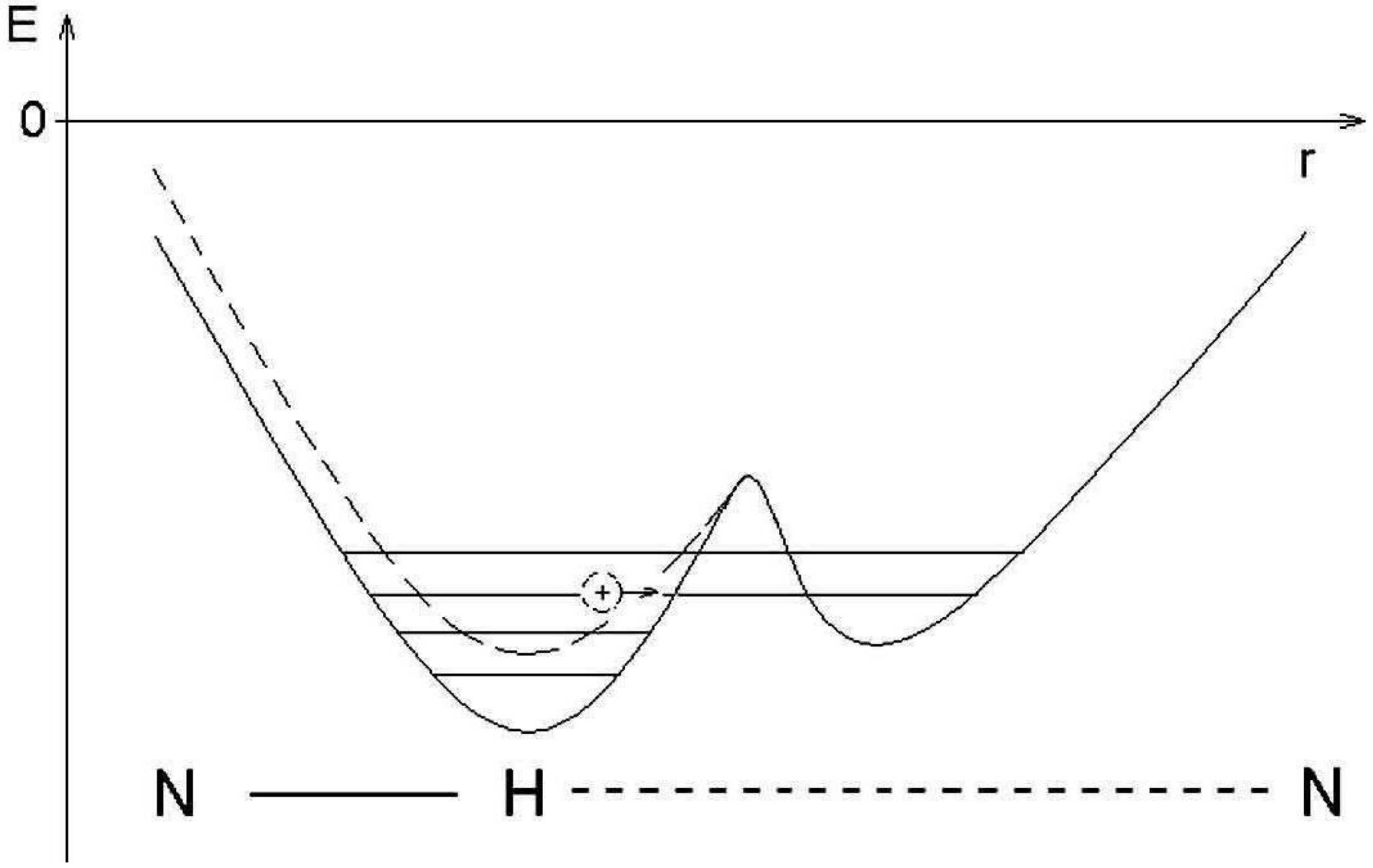

**Figure 32**

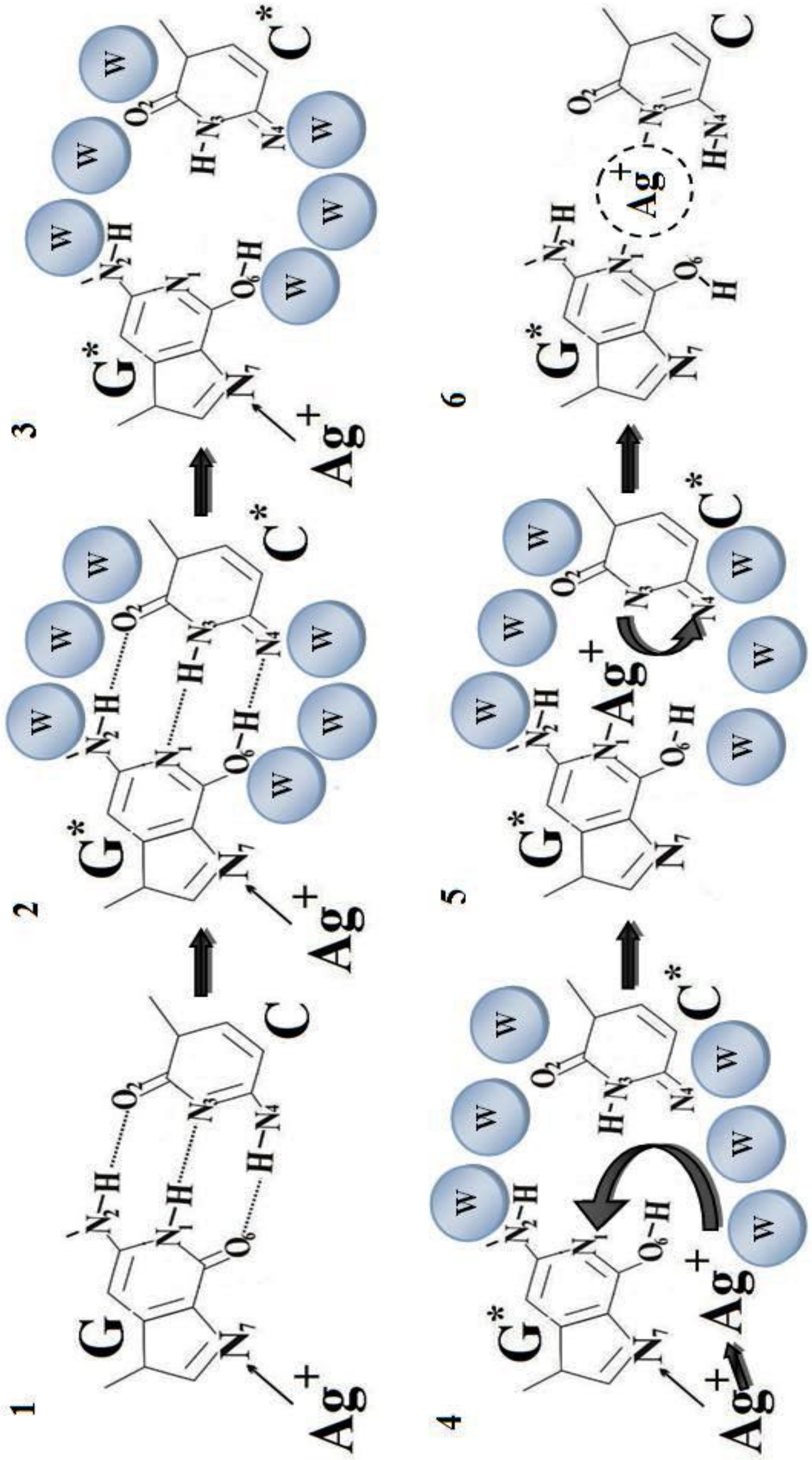

**Figure 33**

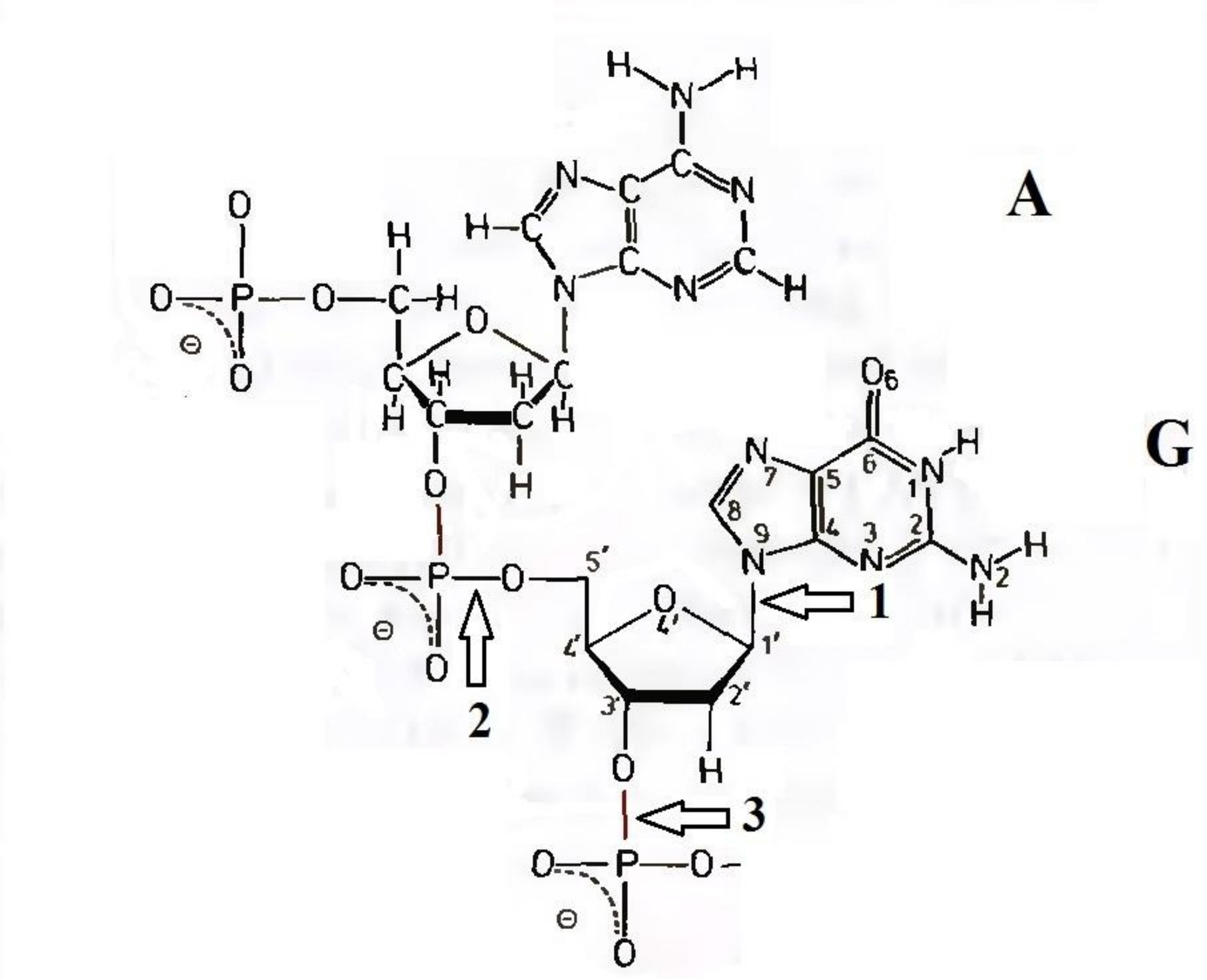

**Figure 34**

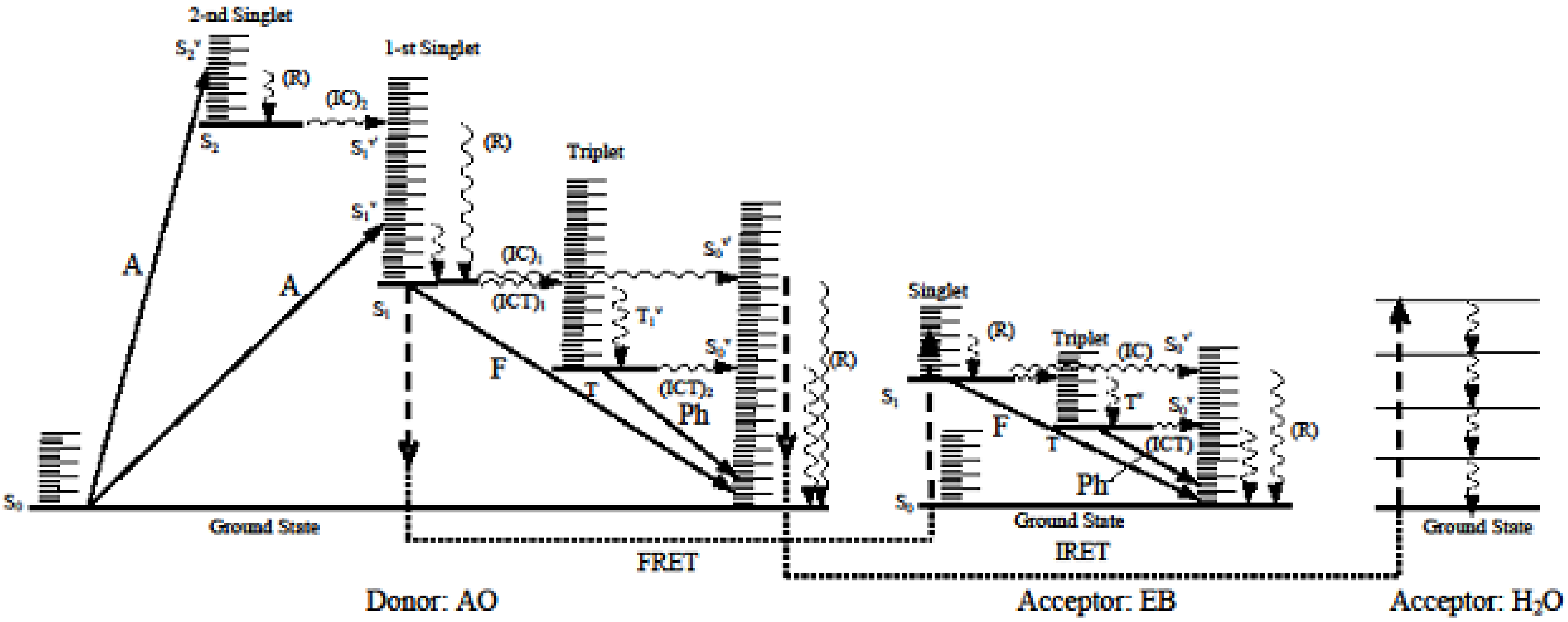

| Designation | Photophysical process | Rate, $s^{-1}$ |
|---|---|---|
| A | Photon absorption: instantaneous process | - |
| F | Fluorescence: singlet-singlet photon emission | $\sim 10^{9}$ |
| Ph | Phosphorescence: triplet-singlet photon emission | $10^{-2}$-$10^{3}$ |
| ICT | Intercombination Transition: nonradiative spin inversion ($S \rightarrow T$ or $T \rightarrow S$ change of multiplicity) | $< 10^{-1}$-$10^{2}$ |
| IC | Internal Conversion: nonradiative (thermal) relaxation of molecule to the electron level of same multiplicity (e.g. $S_2 \rightarrow S_1$) | $10^{11}$ |
| R | Rovibron relaxation from excited vibronic level to the zero level of the same (as usual) state (e.g. $S_1^{v4} \rightarrow S_1^{v0}$) | $10^{12}$ - $10^{14}$ |
| FRET | Fluorescence Resonance Energy Transfer | $10^{12}$ - $10^{13}$ |
| IRET | Inductive resonance energy transfer from excited intercalator to water molecules with resonance overtone vibration of OH. | $10^{12}$ - $10^{14}$ |